\documentclass[reprint, amsmath, amssymb, aps]{revtex4-2}

\UseRawInputEncoding
\usepackage{array}
\usepackage{booktabs}  
\usepackage{graphicx}
\usepackage{dcolumn}
\usepackage{bm}
\usepackage{gensymb}
\usepackage{textcomp}
\usepackage{comment}
\usepackage{xcolor}
\usepackage{hyperref}

\begin{document}
 
\preprint{APS/123-QED}

\title{Quantum Embedding of Non-local Quantum Many-body Interactions in Prototypal Anti-tumor Vaccine Metalloprotein on Near Term Quantum Computing Hardware}

\author{Elena Chachkarova}
\affiliation{King's College London, Theory and Simulation of Condensed Matter
(TSCM), The Strand, London WC2R 2LS, UK, Email:  elena.chachkarova@kcl.ac.uk}

\author{Terence Tse}
\affiliation{King's College London, Theory and Simulation of Condensed Matter
(TSCM), The Strand, London WC2R 2LS, UK, Email:  wai\_hei\_terence.tse@kcl.ac.uk}

\author{Yordan Yordanov}
\affiliation{Alumni - Cambridge University, Cavendish Laboratory, Cambridge CB3 0HE, UK, Email: yy387@cam.ac.uk}

\author{Yao Wei}
\affiliation{King's College London, Theory and Simulation of Condensed Matter
(TSCM), The Strand, London WC2R 2LS, UK, Email: yao.wei@kcl.ac.uk}

\author{Cedric Weber}
\affiliation{King's College London, Theory and Simulation of Condensed Matter
(TSCM), The Strand, London WC2R 2LS, UK, Email: cedric.weber@kcl.ac.uk}

\begin{abstract}
The real world obeys quantum physics and quantum computing presents an alternative way to map physical problems to systems that follow the same laws. Such computation fundamentally constitutes a better way to understand the most challenging quantum problems. One such problem is the accurate simulation of highly correlated quantum systems. Due to the high dimensionality of the problem classical computers require considerable computer power to accurately predict material properties, especially when strong electron interactions are present. Still, modern day quantum hardware has many limitations and only allows for modeling of very simple systems. Here we present for the first time a quantum computer model simulation of a complex hemocyanin molecule, which is an important respiratory protein involved in various physiological processes such as oxygen transport and immune defence, and is also used as a key component in therapeutic vaccines for cancer. To better characterise the mechanism by which hemocyanin transports oxygen, variational quantum eigensolver (VQE) based on fermionic excitations and quantum embedding methods is used in the context of dynamic mean field theory to solve Anderson impurity model (AIM). Finally, it is concluded that the magnetic structure of hemocyanin is largely influenced by the many-body correction and that the computational effort for solving correlated electron systems could be substantially reduced with the introduction of quantum computing algorithms. We encourage the use of the Hamiltonian systems presented in this paper as a benchmark for testing quantum computing algorithms efficiency for chemistry applications.

\begin{description}
\item[Keywords]
Quantum Computing; Variational Quantum Eigensolver; Entanglement; Anderson Impurity Model
\end{description}
\end{abstract}

\maketitle

\section{\label{sec:level1}Introduction}

Quantum computing introduced a new approach in solving challenging computational problems that are difficult for classical computers \cite{quantum-computers-general-1, feynman-quantum-computers}. Development in the field of quantum algorithms as well as quantum hardware has captured the interest of many large companies as well as academic researchers. Limitations of modern day quantum hardware referred to as noisy intermediate-scale quantum (NISQ) devices are being tackled at speed with the utilisation of error mitigation techniques \cite{cai2023quantum, error-mitigation-1, error-mitigation-2} and error corrections \cite{error-correction, error-correction-2}. The current progress in the development of robust quantum hardware \cite{qc-fault-tolerance, noisy-qc, noisy-qc-2, hardware-3, hardware-4} could soon lead to quantum algorithms outperforming the best-known classical algorithms for solving difficult problems with large parameter spaces. Such problems include integer factorisation \cite{shor, integer-fact}, financial portfolio optimisation \cite{finance, finance-2} and fraud detection \cite{finance-qc}, simulation of quantum systems \cite{feynman, qc-qc-3, qc-qc-4, qc-qc-5, vqe-carbon-monoxide, quantum-computers-general-1, feynman-quantum-computers} and more. 

Quantum chemistry is a focus area where quantum computing can overcome many limitations of classical algorithms that become restricted due to the high degree of complexity of the underlying physical processes \cite{Quantum-computational-chemistry, qc-qc, qc-qc-2}. For instance, calculating molecular energies to within chemical accuracy requires an exponential scaling of computational resources and large systems are primarily tackled by costly experimental methods. The introduction of robust quantum algorithms to model and solve electron systems could unleash a new era of material discovery and provide insights into puzzling challenges like room temperature superconductivity and strongly correlated materials modelling. Computation that employs quantum principles is the most fundamental way to find material characteristics and allows to map electron states directly onto qubit quantum states which could in many cases be the most optimal tool for understanding these systems. 

\begin{figure}
\includegraphics[width=1\columnwidth]{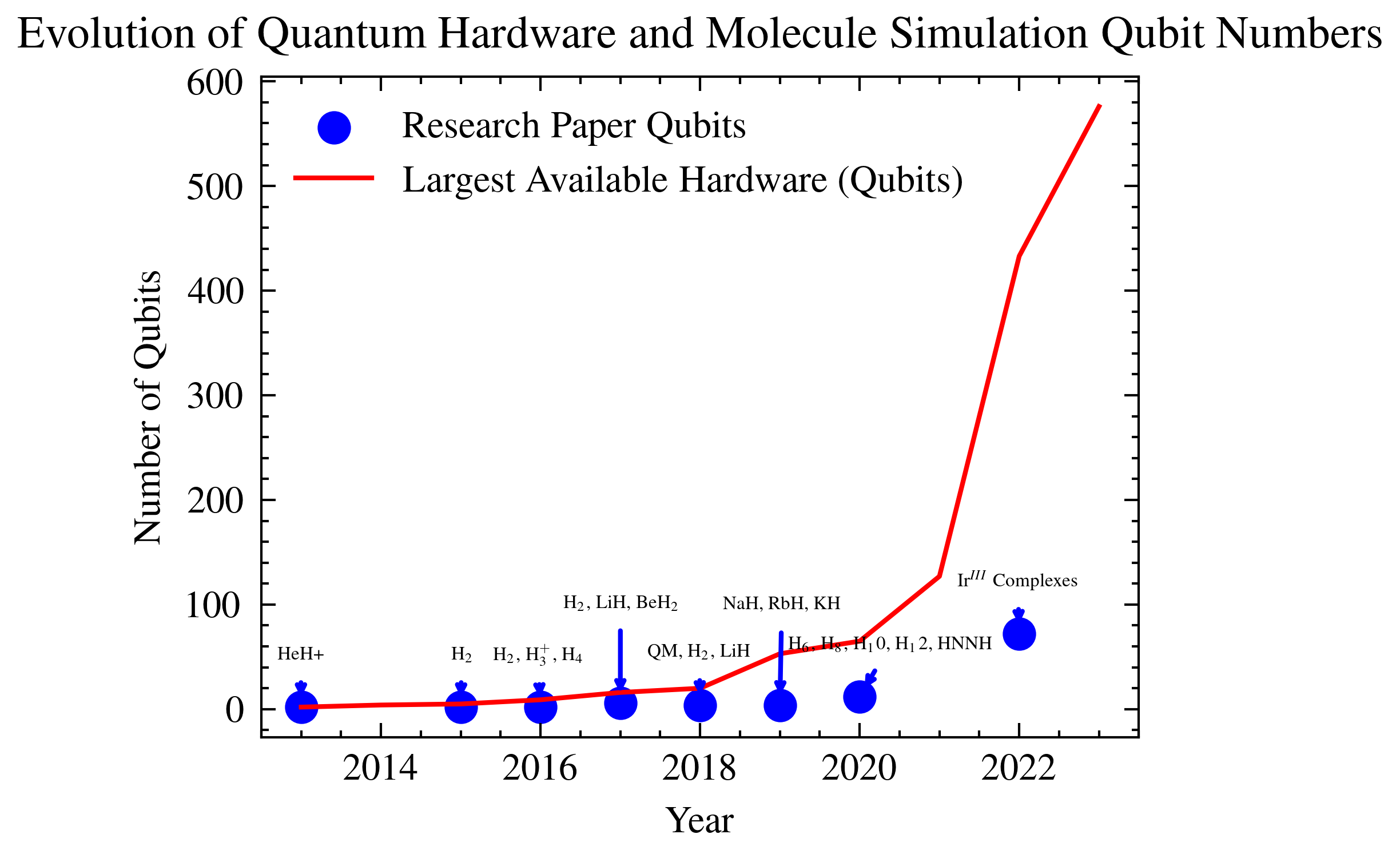}
\caption{Progression of quantum computing technology from 2013 to 2023, highlighting two key trends: the rapid increase in the maximum number of qubits available in quantum hardware (shown in red) and the highest number of qubits used in significant quantum computing based molecule simulation research papers each year (shown in blue). Each point in the research data series is labelled with the specific quantum system studied, providing insight into the scale of quantum experiments and their corresponding hardware capabilities over time. Due to the accumulation of noise, the size of the molecules studied is not increasing as rapid as the advancements in hardware capacity. \cite{Elfving2020HowWQ}}
\label{quantum_experiments_plot}
\end{figure}

So far, quantum hardware has primarily been used to explore the ground state properties of relatively simple molecules with a small number of electrons, such as the water molecule (H$_2$O) \cite{H20, h20-2} and hydrogen molecules \cite{VQE_H, vqe-h}. This focus is due to the increased noise associated with using more qubits and applying entanglement. Recently, there has been growing interest in addressing the challenges associated with strongly correlated electron systems, not only for ground state energy estimation but also for other system characteristics. Notably, work by Microsoft Azure Quantum has introduced an end-to-end comprehensive method for simulating strongly correlated molecular systems. This approach leverages a hybrid classical-quantum method and incorporates the classical shadows technique \cite{classical-shadow} for property measurement, all hosted on Microsoft Azure platform \cite{AzureQuantum}. This paper complements this work by aiming to measure a complex system and its properties using quantum hardware. To expand the context, Fig. \ref{quantum_experiments_plot} illustrates the evolution of quantum computer-based materials modeling research across various molecules of different sizes and complexities, based on data from Table \ref{tab:quantum_experiments}. To estimate the ground state energy of these molecules the variational quantum eigensolver (VQE) method has been employed \cite{VQE-1, VQE-2, VQE-3, vqe-4, VQE-IsingModel}. VQE is a hybrid algorithm that uses both classical computers and quantum computers to estimate the ground state energy of a Hamiltonian. It uses Rayliegh-Ritz variational principle \cite{Ritz} and can model complex ground state wavefunctions in a polynomial time using an ansatz defined by a set of parameters that constitutes a trial measurement which is fed into a classical optimiser that iteratively updates the ansatz parameters to reduce the energy until it converges. The method is sensitive to the form of the ansatz, topology of the hardware, initial state and more, but tailoring those provides flexibility and investigations into VQE simulations can lead to a recipe for the most optimal problem specific VQE setup. Here for the first time we present hardware VQE simulation of the complex hemocyanin molecule \cite{hemocyanin-intro} modelled by Anderson impurity model (AIM) \cite{AIM} on IBM Quantum platform \cite{IBM}. 

The interest in hemocyanin (abbreviated as Hc) came from its fascinating properties and applications, for instance, Hc could be a structural part in metallodrug design but that requires a full understanding of its complex structure which remains challenging to model even with modern day conventional methods \cite{Hemocyanin-Structure}. Hemocyanins are a key component in therapeutic vaccines for cancer due to their useful carrier qualities \cite{MoraRoman2019}. Hemocyanins are also used as nonspecific immunostimulants for the treatment of superficial bladder cancer (SBC), for which these glycoproteins have demonstrated several advantages over more standard immunotherapeutic procedures \cite{Arancibia2012}. Furthermore, the hemocyanin known as keyhole limpet hemocyanin (KLH) has been applied in various in vitro and preclinical studies to determine its effectiveness against other cancers, such as Barrett's adenocarcinoma \cite{McFadden2003}; pancreatic, breast, and prostate cancer \cite{Riggs2005a}; and melanoma \cite{Somasundar2005, Riggs2002}. In nature, Hc is a protein that transports oxygen in the blood of some invertebrate animals and is more resilient to the surrounding environment than haemoglobin. Even though it is less efficient, Hc can be fully functional in low oxygen environments and cold as well as hot temperatures of up to 90 \degree C. Modelling the formation of the oxygenated state of Hc remains a challenge as the binding of the O$_2$ is a spin forbidden transition. To model this system a classic DMFT model \cite{DMFT-1, DMFT-molecule-1, DMFT-molecule-2} would be insufficient as it would treat the Cu atoms separately, and it is believed there is a superexchange pathway between the Cu d-orbitals and intermediate O p-orbitals. Hence, a multi-site AIM is needed to probe the correlated sites directly \cite{DMFT-QC, DFT+DMFT}.

\begin{figure}
\includegraphics[width=1\columnwidth]{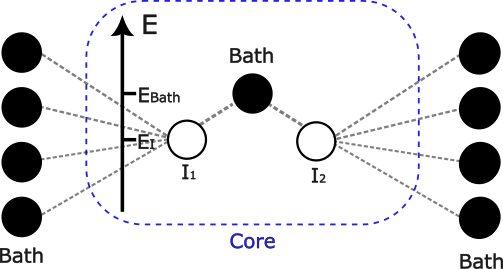}
\caption{Diagram illustrating the topology of the Cu$_2$O model, showing the difference in energy level between bath and impurity sites in the core. Here, the impurities and bath represent the Cu and O atom sites, respectively}
\label{diagram_sep}
\end{figure}

There has been work done for the Cu$_2$O cluster model as shown in the diagram in Fig. \ref{diagram_sep}. It should be noted that the diagram is for illustrative purposes only, and the topology shown can be extended to all-to-all connected models. It can be derived that the Heisenberg antiferromagnetic coupling $J$ can be expressed as 

\begin{equation}
J=4 \frac{t_{p d}^4}{\Delta^2}\left[\frac{1}{U_d}+\frac{1}{\Delta+U_p / 2}\right]
\end{equation}

Where $\Delta$, $t_{pd}$, $U_p$, and $U_d$ represent the charge-transfer energy, Cu-O hopping parameter, and the on-site Coulomb energies at the O and Cu sites respectively \cite{Sheshadri}. Hence, we already have an example of an effective model obtained via embedding, as the bath sites here represent the O$_2$ bridge of hemocyanin rather than the oxygen sites of Cu$_2$O. While most molecular systems are well-captured by DFT (+U) methods \cite{Marzari}, some generate large collections of quantum states and, therefore, would be captured better through quantum computing. One such case is transition metal proteins at certain temperatures, where extensive collections of near-degenerate states exist. These usually occur in magnetic systems near a low-to-high spin transition driven by Hund's coupling. Another case are systems with non-local exchange, such as the superexchange across bridges of transition metal centers. For the former case, work has been done to investigate such systems using quantum methods with the  notable example of iron porphyrin \cite{Cedric_hemoglobin}.

Readers are strongly encouraged to read the “Methods” section before the “Results” section to understand the transformation of a molecular Hamiltonian into a modelled AIM Hamiltonian and the construction of VQE simulation and its components. The “Results” section presents the outcomes of the VQE simulation runs for different VQE setups. The “Discussion” section expands on the meaning of the results and proposes ideas to further the research.

\begin{table*}[ht]
\centering
\label{tab:quantum_experiments}
\textbf{Evolution of Quantum Hardware and Molecule Simulations}
\resizebox{\textwidth}{!}{
\begin{tabular}{@{}lllllll@{}}
\toprule
Year & Qubits & Systems & Methods & Platform & Hardware Max Qubits & Company \\
\midrule
2013 & 2 & HeH+ & VQE-UCC & Silicon Photonic & 2 & In-house \cite{Peruzzo2014} \\
2015 & 2 & HeH+ & VQE-UCC & Trapped ion & 2 & IonQ \cite{Shen2017} \\
2015 & 2 & H$_2$ & VQE-UCC & Superconducting & 9 & Google \cite{OMalley2016} \\
2016 & 2 & H$_2$, H$_3^+$, H$_4$ & IPEA, VQE-UCC & Silicon photonic & 2 & In-house \cite{Santagati2018} \\
2017 & 6 & H$_2$, LiH, BeH$_2$, Heisenberg model & Hardware-efficient VQE & Superconducting & 20 & IBM \cite{Kandala2017} \\
2017 & 2 & H$_2$ (excited states) & Hardware-specific VQE & Superconducting & 20 & UC Berkeley \cite{Colless2018} \\
2018 & 3 & H$_2$, LiH & VQE-UCC & Trapped-ion & 11 & Honeywell \cite{Hempel2018} \\
2018 & 4 & Quantum magnetism, H$_2$, LiH & Hardware-efficient VQE & Superconducting & 20 & IBM \cite{Kandala2019} \\
2018 & 4 & H$_2$, LiH & Qubit CC & Superconducting & 53 & Google \cite{Arute2020HF} \\
2019 & 2 & H$_2$O & QPE & NMR & 2 & In-house \cite{Li2019} \\
2019 & 4 & H$_2$O & VQE-UCC & Trapped-ion & 11 & IonQ \cite{Nam2020} \\
2019 & 4 & NaH, RbH, KH & Hardware-efficient VQE(-UCC) & Superconducting & 53 & ORNL \cite{McCaskey2019} \\
2019 & 2 & Lithium superoxide dimer & VQE-UCC & Superconducting & 53 & IBM \cite{Gao2019} \\
2019 & 3 & H$_3$ & VQE-UCC & Superconducting & 53 & Google \cite{Smart2019} \\
2020 & 12 & H$_6$, H$_8$, H$_10$, H$_12$, HNNH & VQE-HF & Superconducting & 53 & Google \cite{Arute2020HF} \\
2020 & 2 & PSPCz, 2F-PSPCz, 4F-PSPCz & qEOM-VQE, VQD & Superconducting & 127 & IBM \cite{Gao2020} \\
2022 & 28 & C$_2$H$_4$ & Point Symmetry & Emulator & 433 & Quantum Emulation Group \cite{28_qubit} \\
2022 & 72 & Ir$^{III}$ Complexes & Point Symmetry & Emulator & 1000+ & Quantum Emulation Group \cite{72_qubit} \\
\textcolor{blue}{2024} & \textcolor{blue}{6} & \textcolor{blue}{oxyHc (58-atom model)} & \textcolor{blue}{VQE-UCC} & \textcolor{blue}{Superconducting} & \textcolor{blue}{7} & \textcolor{blue}{IBM \cite{IBM}} \\
\textcolor{blue}{2024} & \textcolor{blue}{6} & \textcolor{blue}{oxyHc (58-atom model)} & \textcolor{blue}{VQE-UCC} & \textcolor{blue}{Superconducting} & \textcolor{blue}{20} & \textcolor{blue}{Quantinuum \cite{quantinuum}} \\
\textcolor{blue}{2024} & \textcolor{blue}{14} & \textcolor{blue}{oxyHc (58-atom model)} & \textcolor{blue}{VQE-UCC} & \textcolor{blue}{Emulator} & \textcolor{blue}{50+} & \textcolor{blue}{IBM \cite{IBM}} \\
\bottomrule
\end{tabular}
}
\caption{Quantum Experiments Summary - Evolution of quantum chemistry molecule simulations on quantum hardware and emulators from 2013 to the present. The recent entries, highlighted in the table's final two rows in blue, demonstrate a consistency in the size of qubits employed but also an expansion to encompass larger molecular systems. \cite{Elfving2020HowWQ}.}
\end{table*}

\section{Methods} 

The following investigation utilised a multi-site Anderson impurity model (AIM) of the Hamiltonian of Hc molecule developed previously in \cite{Cedric-hemocyanin} to measure its properties on IBM \cite{IBM} and Quantinuum \cite{quantinuum} quantum machines with different number of qubits and VQE solver setups. The subsections below identify the steps to obtain the model Hamiltonian and describe the VQE algorithm setup. \\

\noindent{{\large Molecular Modelling}}

\begin{figure}
\includegraphics[width=1\columnwidth]{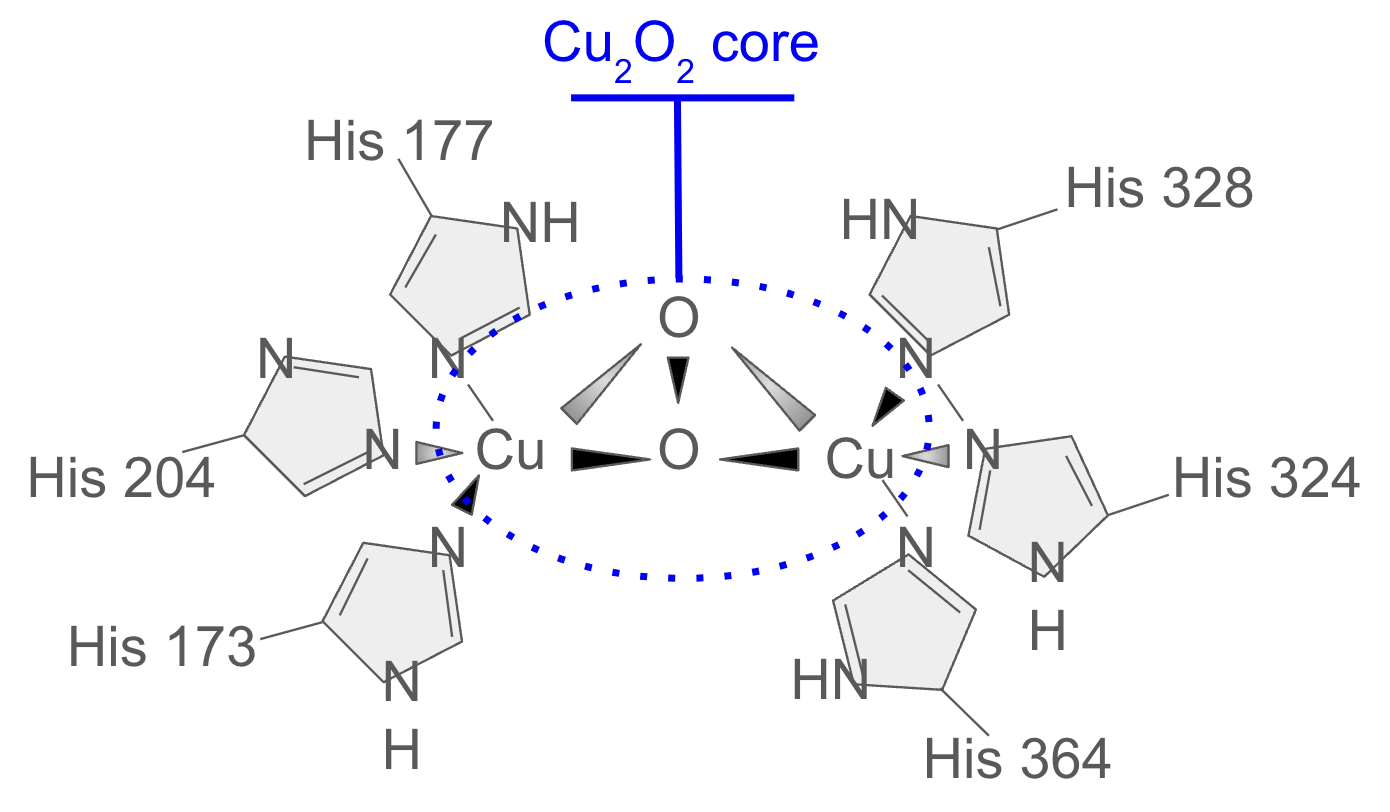}
\caption{Hemocyanin (oxyHc) $O_2$-bound form with active site $Cu_2O_2$ core labelled in blue (the $Cu_2$ center is a dication, charge not shown).}
\label{hc}
\end{figure}

\noindent{Due to the complex character of the binding of oxygen in the formation of oxygenated hemocyanin (oxyHc diagram in Fig. ~\ref{hc}), most theoretical studies have failed to correctly predict its properties. Here, we follow steps from a previous study on Hc \cite{Cedric-hemocyanin} that has shown agreement with experimental results. The work presents a density functional theory (DFT) + dynamical mean field theory (DMFT) simulation on a 58-atom model of the oxyHc. To accurately characterise the superexchange mechanism between the Cu$_2$ d-orbitals and intermediate O p-orbitals, a non-local (cluster) DMFT is employed. A multi-site Anderson impurity model (AIM) captures all correlated sites which labels the method better as DFT+AIM. A self-consistency cycle over the charge density is performed to ensure a fixed number of electrons.}

Anderson impurity model:
To solve with cluster DMFT \cite{Cluster-DMFT}, we map the hemocyanin core onto an Anderson-impurity model. 

The AIM can be defined as:

\begin{multline}
H_{AIM}=\sum_{mn\sigma}\left(\epsilon^n_{mn\sigma}a^{\dagger}_{m\sigma}a_{n\sigma}+\epsilon^a_{mn\sigma}a^{\dagger}_{m\sigma}a^{\dagger}_{n-\sigma}+h.c.\right)+ \\
\sum_{mi\sigma}V_{mi\sigma}\left(a^{\dagger}_{m\sigma}c_{i\sigma}+h.c.\right)+\mu\sum_{i\sigma}c^{\dagger}_{i\sigma}c_{i\sigma}+\sum_{i}U\hat{n_{i\downarrow}}\hat{n}_{i\uparrow}
\end{multline}

The fermionic operators a$^{\dagger}_{mn}$ (a$_{mn}$) create (destroy) a particle in the bath, and the fermionic operators c$^{\dagger}_i$(c$_i$) create (destroy) a particle in the cluster of impurities. The indices m, n run over the bath sites, and the index i is running over the impurity sites, the sites of the bath are connected by long-range hopping matrix elements through the particle-hole (particle-particle) channel $\epsilon^{n}$ ($\epsilon^{a}$ ), the non-correlated sites of the bath are also connected to the correlated impurities by the matrix elements V$_{mi}$, the onsite repulsion at the impurity sites is U, and $\mu$ is the chemical potential.

Here, we map the hemocyanin core onto the AIM by taking the two copper sites as the two impurity sites, with the dioxygen bridge and imidazole ligands acting as the bath. 

DFT in conjunction with DMFT is used to retrieve the AIM parameters. Non-spin polarised DFT is carried out with PBE functionals, with Non-orthogonal Generalized Wannier Function (NGWF) \cite{Wannier} gradient threshold of $2*10{^-6}$ \cite{PBE}. A Hubbard U correction of 10 eV is applied to all d-orbitals.

DMFT is carried out with U=0.29340 E$_h$ and J$_{hund}$=0.02939, using the Lanczos
impurity solver over 2000 Matsubara frequencies and 800000 conjugate gradient steps \cite{Lanczos}.

As such, we obtain the interaction parameters and on-site energies as presented in the Supplementary information. \\

\noindent{{\large Variational Quantum Eigensolver}}

\noindent{VQE is based on the Rayliegh-Ritz variational principle \cite{Ritz} which finds an upper bound estimate of the expectation value of an observable through the application of a trial wavefunction. In the context of Quantum Chemistry this accounts for the optimisation of ground state energy $E_0$ of a system with Hamiltonian $\hat{H}$ and a trial wavefunction $|\psi_{trial} \rangle$, as follows}

\begin{eqnarray}
E_{0}\leqslant\frac{\langle \psi_{trial}|\hat{H}|\psi_{trial} \rangle}{ \langle \psi_{trial}|\psi_{trial} \rangle}.
\label{initialVQE}
\end{eqnarray}

A main part of constructing a robust VQE simulation is choosing a trial wavefunction with an optimal parametrisation that can bring the value of the ground state energy estimate to the exact energy within the desired level of precision. In quantum computation, the parametrisation step has been in focus of much research due to the vast number of possibilities from hardware efficient to problem-tailored trial wavefunction, also called ansatz wavefunction. Building an ansatz on a quantum device constitutes the application of quantum gates which are represented by parameterised unitary operators, labelled $U(\theta)$ applied to $N$ qubits, usually initialised in state $|0>^{\bigotimes N}$, where $\theta$ is a set of parameters that take values in the scope $(-\pi,\pi ]$.

To represent the Hamiltonian in a form directly measurable on a quantum device it is transformed into a weighted sum of spin operators using a common mapping: Jordan–Wigner transformation (Jordan and Wigner 1928) \cite{JW}. Qiskit does not provide an out-of-the-box mapper between OpenFermion \cite{openfermion} fermionic Hamiltonian and qiskit Pauli strings representation \cite{qiskit-book} as a feature for bespoke Hamiltonians, hence we implemented that step separately. The Hamiltonian is then defined using Pauli strings of the form $\hat{P_a} \in \{ I, X, Y, Z \}^{\bigotimes N}$, with $N$ the number of qubits used to model the wavefunction,

\begin{eqnarray}
\hat{H}=\sum_{a}^{S}w_{a}\hat{P_{a}}.
\end{eqnarray}

\noindent{With this representation Eq.~\ref{initialVQE} using the AIM Hamiltonian, $H_{AIM}$, can be transformed into}

\begin{eqnarray}
E_{\mathrm{VQE}}=\underset{\theta}{\min}\sum_{a}^{S}w_{a}<0|U^{\dagger}(\theta)\hat{P}_{a}U(\theta)|0>,
\end{eqnarray}

\begin{figure}
\includegraphics[width=1\columnwidth]{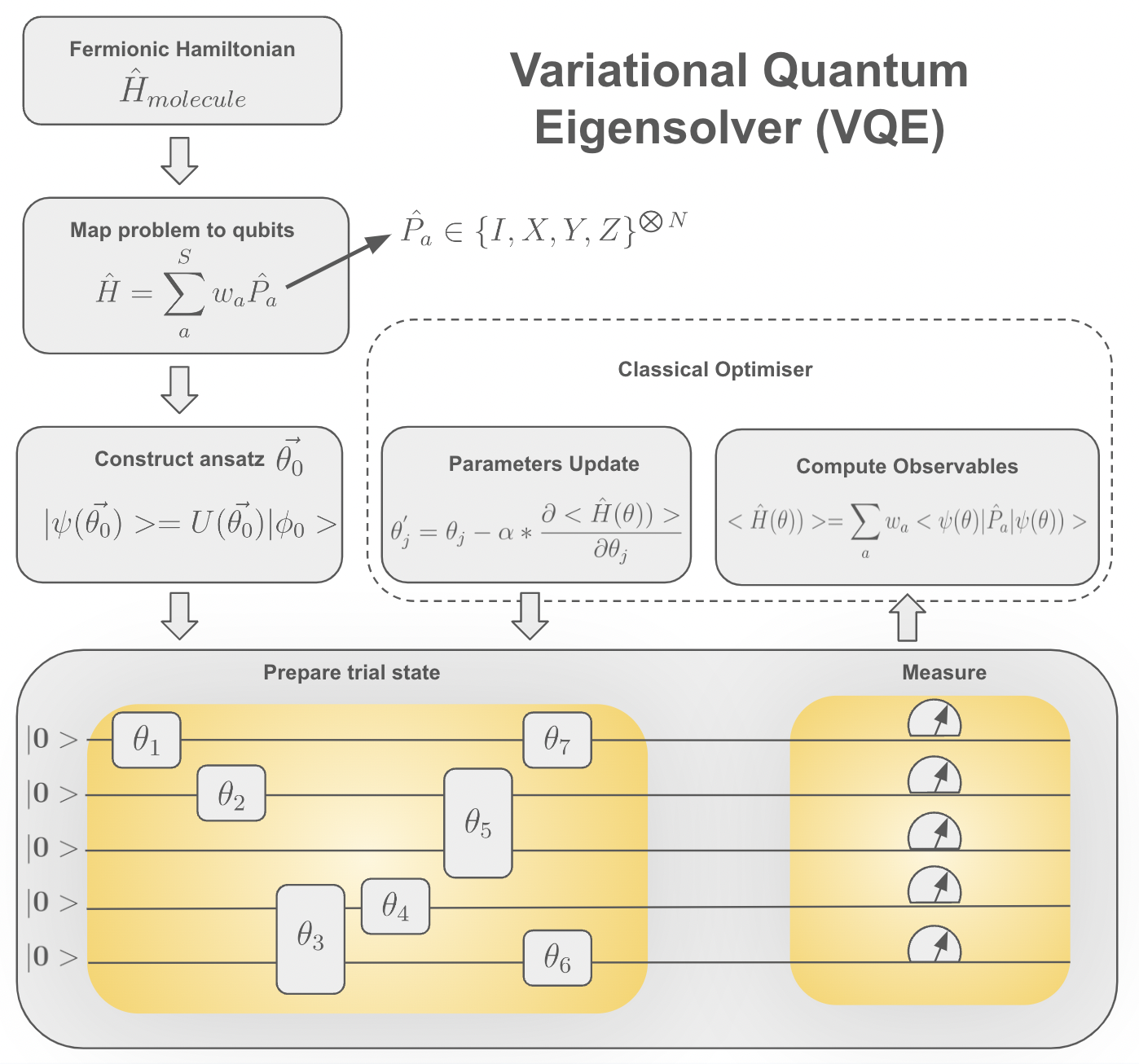}
\caption{Schematic representation of the variational quantum eigensolver (VQE) method for minimisation of a molecular Hamiltonian by adjusting variational parameters, $\vec{\theta}$. The simulation starts by retrieving the molecular Hamiltonian and mapping it onto a qubit Hamiltonian. Then the chosen ansatz is applied to the initialised qubit register. After measurements and computation of the observables new values for the $\vec{\theta'}$ are extracted and fed back to the beginning of the loop.}
\label{vqe}
\end{figure}

\noindent{where each Pauli string term $<0|U^{\dagger}(\theta)\hat{P}_{a}U(\theta)|0>$ is measured on a quantum device whilst the summation and minimisation steps are performed on a conventional computer using a classical optimiser, hence the hybrid nature of the quantum computer based VQE simulations. The full VQE cycle can be seen in Fig. ~\ref{vqe}.}

The design and mapping of the VQE simulation are of great importance to the quality of the results on modern day NISQ devices, for instance, the depth of the circuits and the number of CNOT gates strongly affect the quantum noise on the measurements with shorter circuits generally leading to more accurate results. The choice of the ansatz is possibly the most important step of the simulation, as shown in Sec. ~\ref{section3}. There are three main categories of ansatz types and their performance is strongly dependent on the problem in hand. These include the Unitary Coupled Cluster (UCC) type ansatz \cite{ucc-1, ucc-2}, the Hardware-Efficient Ansatz (HEA) \cite{hea-1, hea-2, hea-3, hea-4, hea-5} and a third type which is a mixture between UCC and HEA. 

The unitary coupled cluster with singles and double excitations (UCCSD) ansatz \cite{uccsd-1, uccsd-2, uccsd-3} is a chemistry-inspired ansatz that incorporates knowledge of the underlying quantum system with terms in the ansatz representing specific electronic configurations. The UCCSD trial state is prepared from a reference state $|\phi_0>$ by applying exponentiated excitation operators, with $|\phi_0>$ commonly chosen as a Hartree–Fock mean-field wave function.

\begin{eqnarray}
\hat{T}_1=\sum_{i=1}^{N_{occ}}\sum_{a=1}^{N_{virt}}t_{i}^{a}\hat{a}_{a}^{\dagger}\hat{a}_{i}
\end{eqnarray}
\begin{eqnarray}
\hat{T_2}=\sum_{i,j=1}^{N_{occ}}\sum_{a,b=1}^{N_{virt}}t_{ij}^{ab}\hat{a}_{a}^{\dagger}\hat{a}_{b}^{\dagger}\hat{a}_{j}\hat{a}_{i}
\end{eqnarray}

\noindent{Here, $\hat{T_1}$ and $\hat{T_2}$ are the single and double excitation terms with $\hat{a^\dagger}$ and $\hat{a}$ being the creation and annihilation operators, $t_{i}^a$ are the single excitation amplitudes and $t_{ij}^{ab}$ are the double excitation amplitudes. $N_{occ}$ and $N_{virt}$ represent the number of occupied and virtual orbitals, respectively. For UCCS ansatz only $\hat{T_1}$ terms are used, whilst for UCCSD both terms are applied. A second degree truncation of the excitations (UCCSD case) forms the following trial wavefunction.}

\begin{eqnarray}
\hat{T}=\hat{T_1}+\hat{T_2}
\end{eqnarray}
\begin{eqnarray}
U(\vec{\theta})=e^{\hat{T}-\hat{T}^\dagger}
\end{eqnarray}
\begin{eqnarray}
|\psi_{trial}>= |\psi(\vec{\theta})> = e^{\hat{T}-\hat{T}^\dagger}|\phi_0>
\end{eqnarray}

\begin{figure}
\includegraphics[width=1\columnwidth]{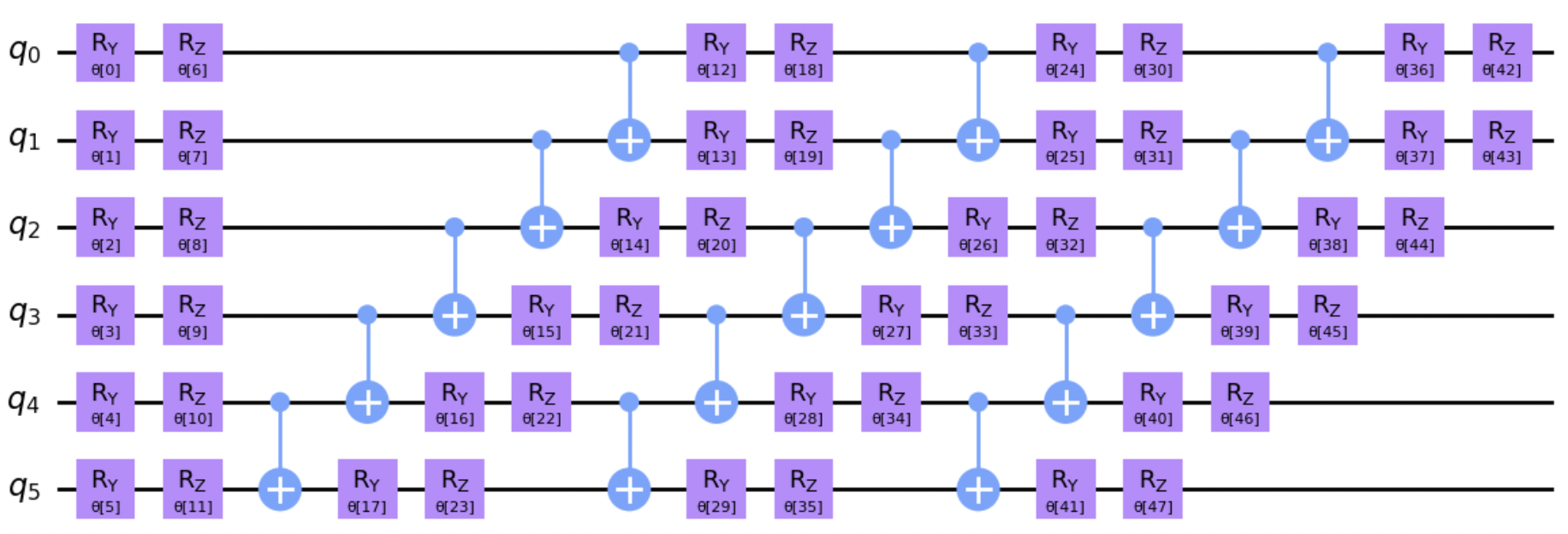}
\caption{Diagram of EfficientSU2 circuit \cite{Nakanishi2020} consisting of layers of single qubit operations spanned by SU(2) and CX entanglements \cite{IBM}. Hardware efficient ansatz applied on 6 qubits.}
\label{hardware_ansatz}
\end{figure}

\noindent{The other type of trial wavefunctions is the hardware-efficient ansatz. These ansatzes are constructed from a limited set of gates that are easy to implement on quantum hardware, but have no chemical interpretation. An example one is the EfficientSU2 ansatz \cite{Nakanishi2020}, which is used in this investigation, Fig. ~\ref{hardware_ansatz}.}

In this work, we compare the accuracy of VQE simulations on modelled hemocyanin with UCCS, UCCSD, HEA and bespoke ansatzes. Qiskit implementation allows for adjusting of the ansatz parameters, such as specifying the set of excitations or selecting spin conserving excitations only. The variations in the results between these are presented in the next section.

\section{Results}
\label{section3}

\noindent{{\large Ground state energy measurement}}

\begin{figure}
\includegraphics[width=1\columnwidth]{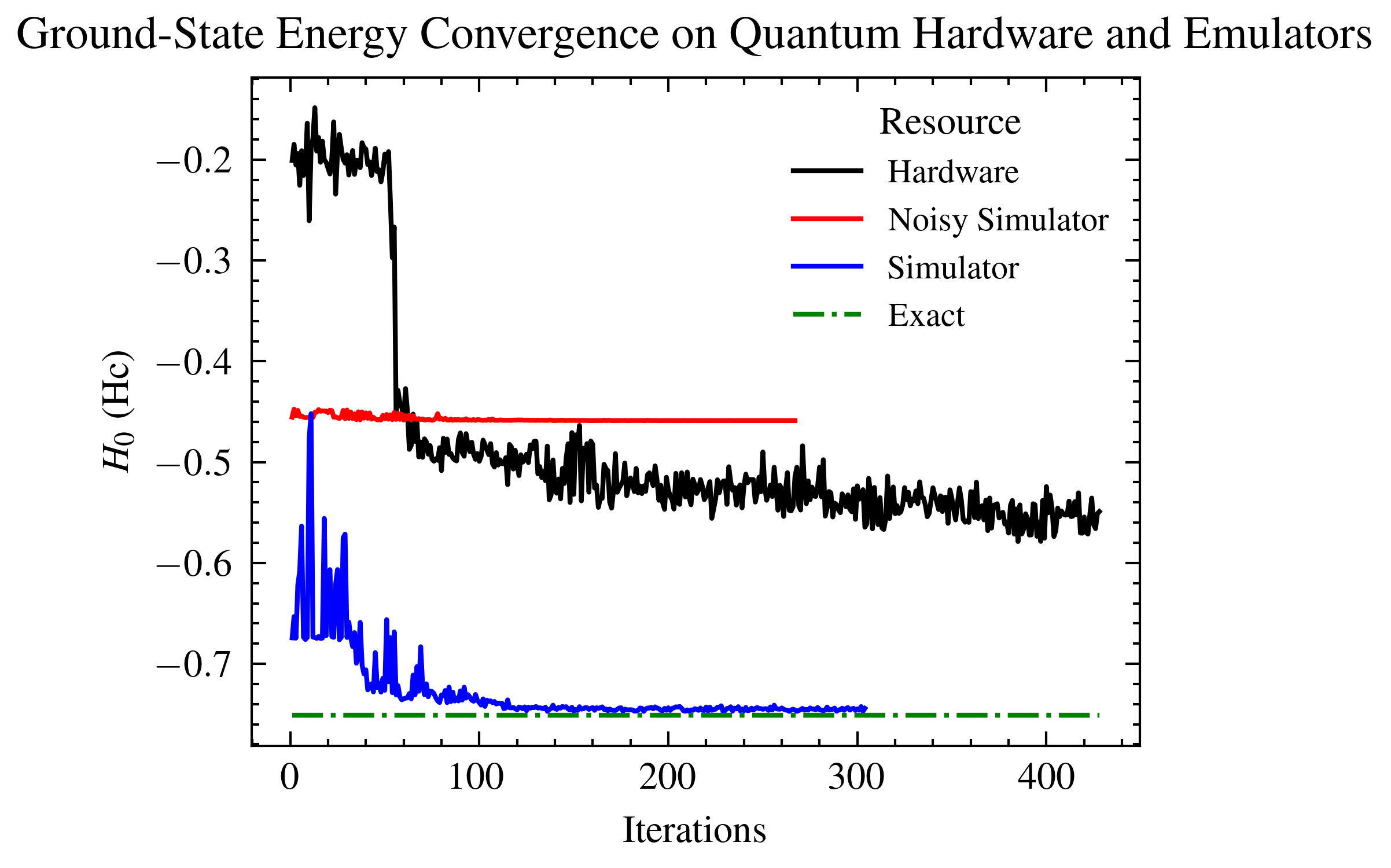}
\caption{Ground-state energy estimates  in $E_h$ for a 6-qubit hemocyanin model using VQE simulations on both the IBM Perth hardware and the IBM QASM simulator. Results from the noiseless simulator converge to the expected outcome within 120 iterations, demonstrating high accuracy. Conversely, the hardware execution reaches a barren plateau \cite{barren}, exhibiting only minimal, continuous decline. Additionally, the noisy simulator, employing a noise model replicated from the actual hardware, fails to identify any feasible solution, attributable to the absence of error mitigation and correction techniques. In each instance, the setup employed a Generalized UCCSD ansatz.}
\label{ground_energy_hemocyanin}
\end{figure}

\begin{figure}
\includegraphics[width=1\columnwidth]{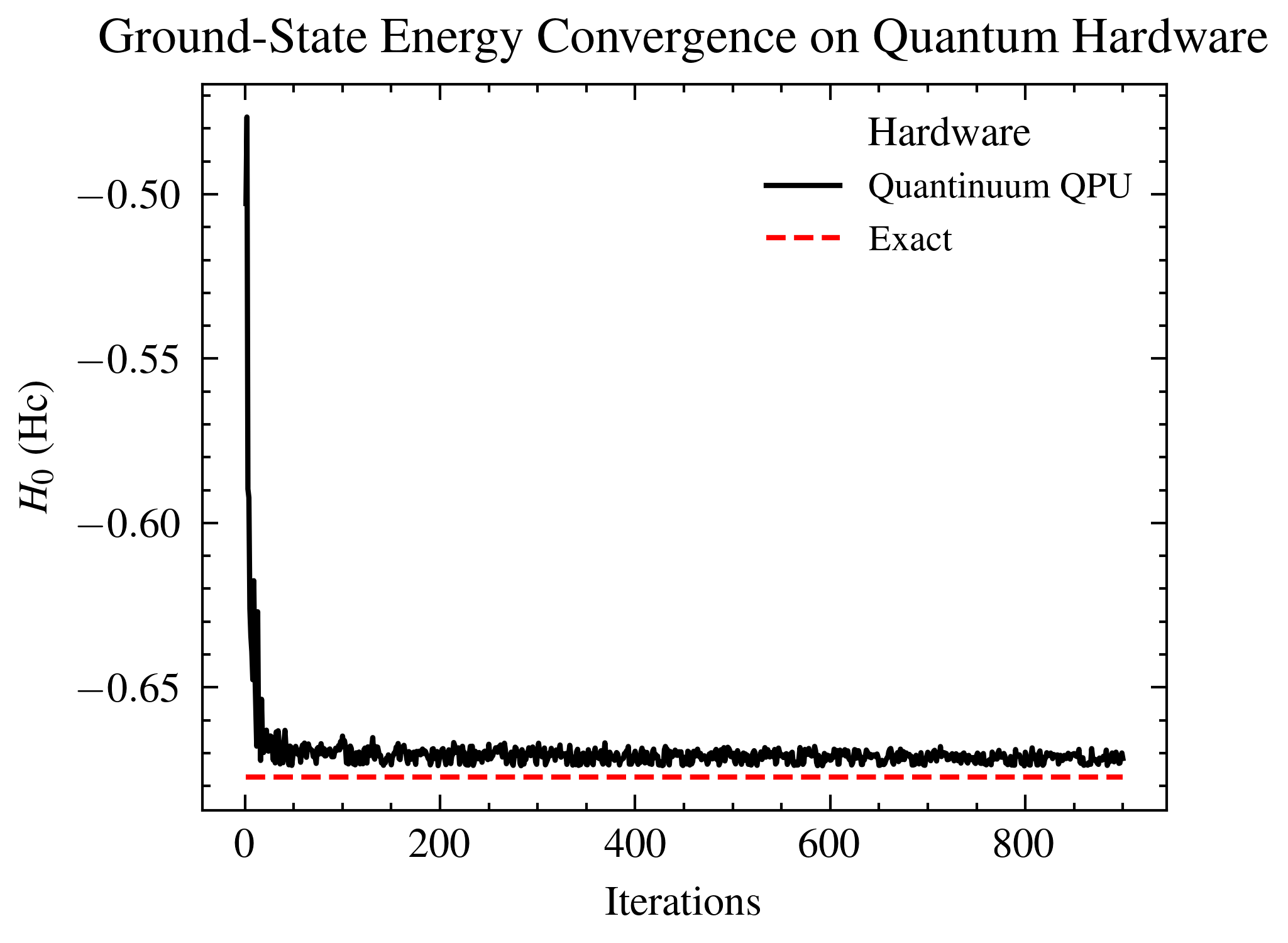}
\caption{Ground-state energy estimates in $E_h$ for a 6-qubit hemocyanin model from VQE simulation on Quantinuum hardware using Microsoft Azure Quantum platform \cite{microsoft_azure_quantum}. The VQE simulation employed a generalised UCCSD ansatz and SPSA \cite{spsa} optimiser with maximum iterations set to 10000. The simulation shows a clear convergence to the exact result within accuracy of $\approx0.485\%$ to the exact value.}
\label{quantinnium}
\end{figure}

\noindent{We implemented qiskit VQE algorithm on 6, 8, 10, 14  and 16 qubit hemocyanin AIM model Hamiltonians on IBM quantum simulator with and without noise and on IBM and Quantinuum quantum hardware. Results in the paper present data from the 6 and 14 qubit runs to reduce duplication, to access the extra runs and Hamiltonians consult with the code. In all cases, the simulations employed a generalised UCCSD ansatz. Fig. ~\ref{ground_energy_hemocyanin} shows a comparison between the converged values as well as the exact value. The noisy simulator (noise mapped from ibm\_Casablanca) VQE run did not show convergence without any error mitigation techniques. The noise in this case is too large compared to the changes in the energy estimates from the VQE excitations and the simulation fails to progress. Fig. ~\ref{quantinnium} shows clear convergence on Quantinuum QPU within $0.485\%$ accuracy of the ground state energy estimate.\\}

\noindent{{\large Ansatz construction}}

\begin{figure}
\includegraphics[width=1\columnwidth]{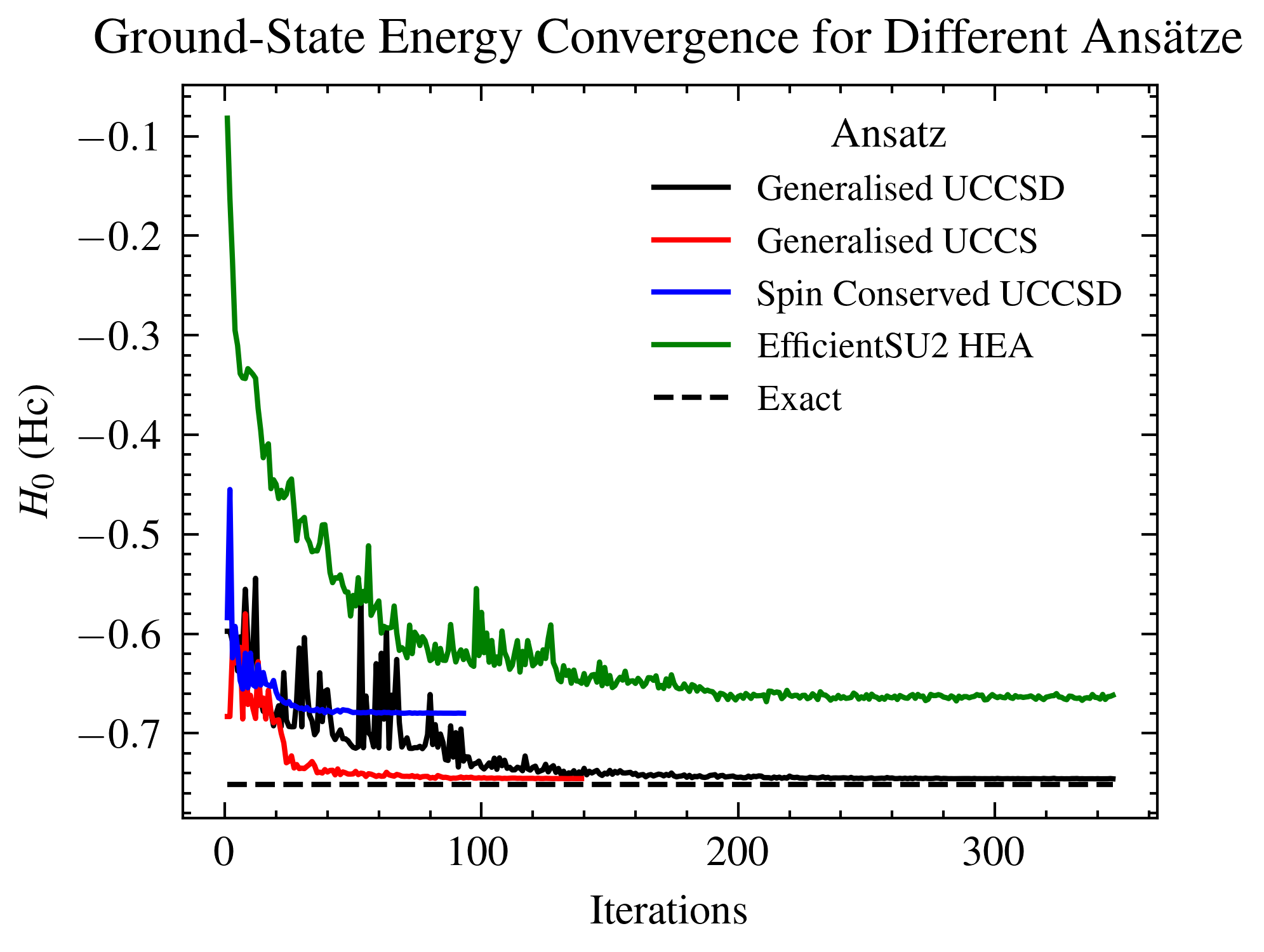}
\caption{Comparison of ground-state energy estimates in $E_h$ derived from the incorporation of additional interactions in the UCC ansatz states, labeled according to ansatz construction parameters. The Generalised UCCS ansatz identifies a feasible solution most rapidly. Additionally, the inclusion of second-order excitations also results in correct convergence. In contrast, the EfficientSU2 HEA ansatz converges to a barren plateau. The spin-conserved UCCSD ansatz achieves convergence swiftly probably due to the restricted number of pool operators, yet it also plateaus.}
\label{ansatz}
\end{figure}

\noindent{Choosing the most optimal ansatz for a VQE simulation is of great importance and can considerably change the final result. In Fig. ~\ref{ansatz} we present a plot of the model Hc Hamiltonian ground state energy estimation on an IBM simulator for different forms of the ansatz: generalised UCC-S/SD type ansatz, spin conserving UCCSD ansatz and hardware efficient ansatz; all compared to the exact value. As expected the output of the simulation runs follow different convergence, which testifies for the strong sensitivity of VQE to the ansatz. \\} 

\noindent{{\large Noise models}}

\begin{figure}
\includegraphics[width=1\columnwidth]{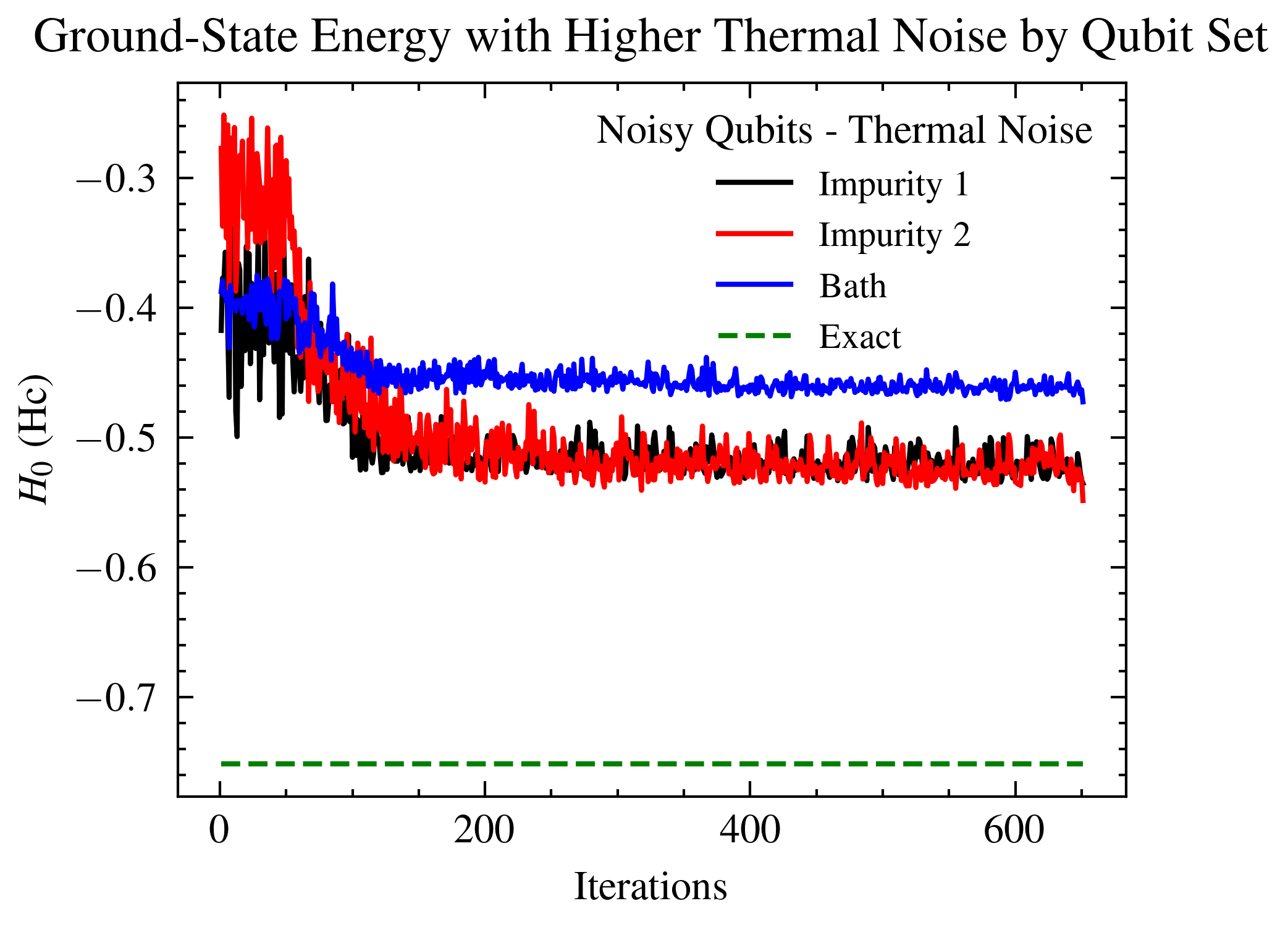}
\caption{Comparison of ground-state energy estimates in $E_h$ from various 6-qubit VQE simulation setups utilizing the Generalised UCCSD ansatz, Hamiltonian $Config - A$, and specific VQE parameters. The simulations were subjected to \(T_1/T_2\) thermal relaxation noise, selectively applied to three different sets of qubits: Impurity \(1\) (qubits \(0\) and \(1\)), Impurity \(2\) (qubits \(2\) and \(3\)), and the Bath (qubits \(4\) and \(5\)). Consistent with expectations due to the correlation between Impurity \(1\) and Impurity \(2\) sites, noise impacting either set results in similar values for the ground state energy. Conversely, noise affecting the Bath sites demonstrates more significant detrimental effects on the system's performance.}
\label{noise_qubits-higher-thermal}
\end{figure}

\begin{figure}
\includegraphics[width=1\columnwidth]{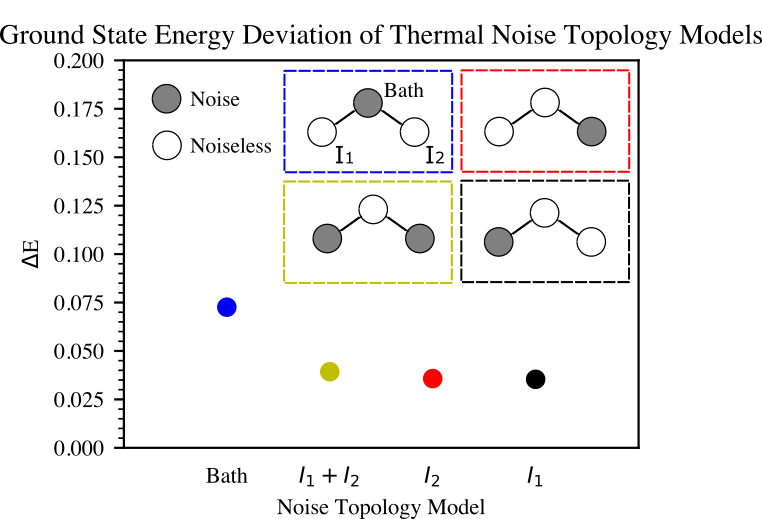}
\caption{Comparison diagram of ground-state energy deviation in $E_h$ from various 6-qubit VQE simulation setups utilizing the Generalised UCCSD ansatz, Hamiltonian $Config - A$, and specific VQE parameters. The simulations were subjected to \(T_1/T_2\) thermal relaxation noise, selectively applied to four different sets of qubits: Impurity \(1\) (qubits \(0\) and \(1\)); Impurity \(2\) (qubits \(2\) and \(3\)); Impurity \(1\) and Impurity \(2\) (qubits \(0\), \(1\), \(2\) and \(3\)); and the Bath (qubits \(4\) and \(5\)). Consistent with expectations due to the correlation between Impurity \(1\) and Impurity \(2\) sites, noise impacting either or even both sets results in similar values for the ground state energy. Conversely, noise affecting the Bath sites demonstrates more significant detrimental effects on the system's performance.}
\label{noise-diagram}
\end{figure}

\begin{figure}
\includegraphics[width=1\columnwidth]{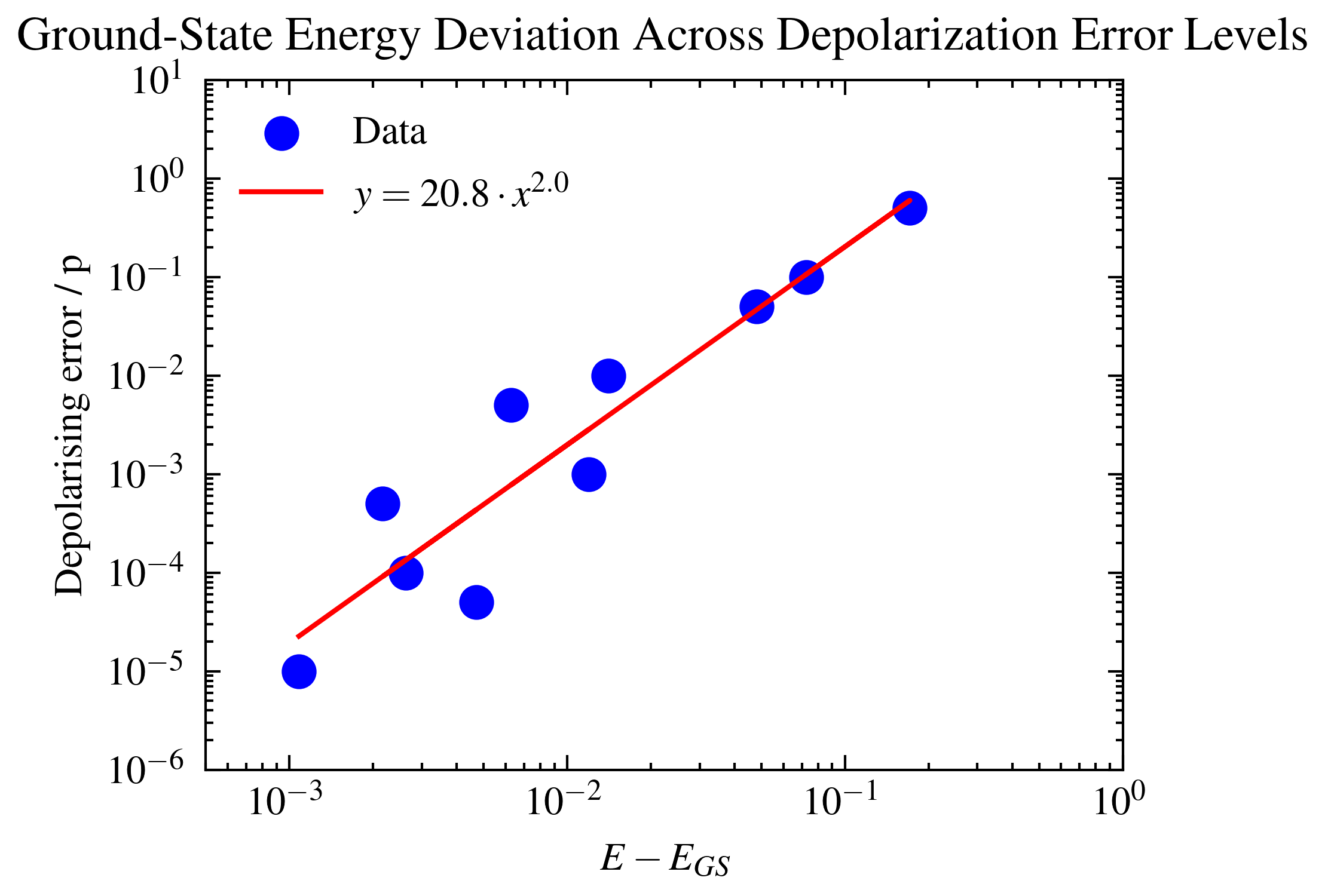}
\caption{Log-log plot illustrating the difference between the exact ground state energy, \(E_{GS}\), in $E_h$ and noisy VQE simulation outcomes across a spectrum of uniform depolarization error rates, ranging from \(10^{-1}\) to \(10^{-5}\). Each simulation was aligned with the coupling map of IBM\_Casablanca to reflect the realistic connectivity between qubits. These ground-state energy estimates were derived from identical 6-qubit VQE simulation configurations employing a Generalised UCCSD ansatz, Hamiltonian Config-A, and specified VQE parameters. The observed linear dependence on the log-log scale indicates a systematic and predictable influence of depolarization errors on simulation accuracy with a slope of approximately $2.01$.}
\label{noise_qubits-depol}
\end{figure}

\begin{figure}
\includegraphics[width=1\columnwidth]{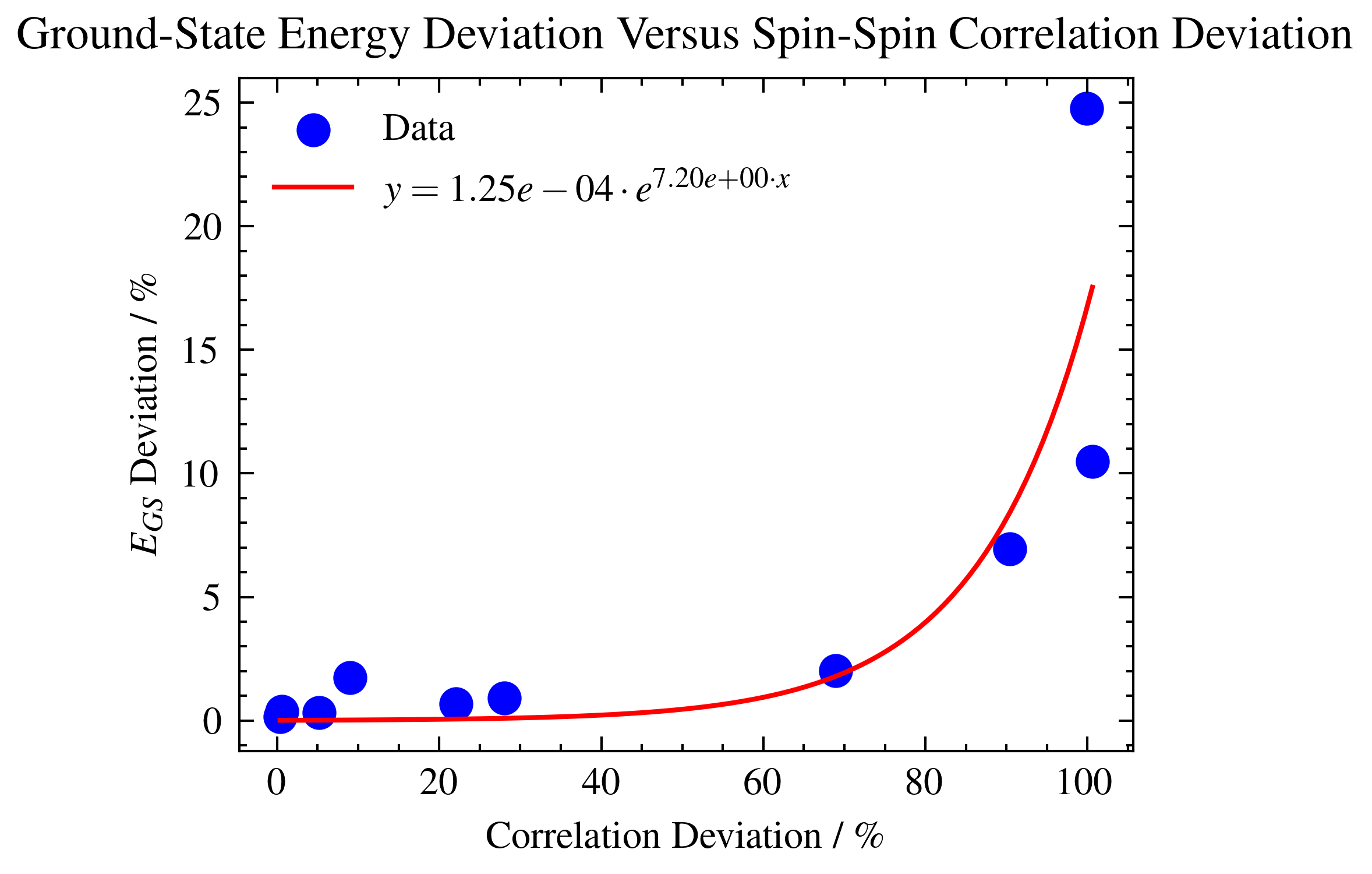}
\caption{Percentage deviations of noiseless to noisy spin-spin correlation (\(\langle S_z[i] * S_z[j] \rangle\)) estimates from VQE simulation outcomes across a spectrum of uniform depolarization error rates, ranging from \(10^{-1}\) to \(10^{-5}\) versus ground state energy, \(E_{GS}\), deviations for the same level of noise. Each simulation was aligned with the coupling map of IBM\_Casablanca to reflect the realistic connectivity between qubits. The simulation configurations employ a Generalised UCCSD ansatz, Hamiltonian Config-A, and specified VQE parameters. The observed nonlinear relationship in the plot suggests that the correlation results are significantly less stable than the ground state energy estimates when the same level of noise is present. Exponential fit is added for reference.}
\label{noise_qubits-depol-corr}
\end{figure}

\noindent{Real quantum devices have many sources of error. Previously, we investigated the effect of mapping the total noise from a real quantum computer onto a simulator. This entails mapping the error rates on instructions which are determined by gate times and qubit $T1$, $T2$ values, where $T1$ is an energy relaxation time (the time taken for the excited $|1⟩$ state to decay toward the ground state $|0⟩$) and $T2$ is a dephasing time constant. Here, we add custom noise to only specific qubits with the aim to understand better which parts of the simulation cause higher instabilities. We applied $T1/T2$ thermal relaxation noise to the qubits that represent impurity $1$, impurity $2$, both impurities sites ($1$ and $2$) and the bath site in turn. The Hamiltonian was mapped into the $Config - A$ configuration, reference Fig. ~\ref{config-draw}, in all four cases. Results from the simulations are in Fig. ~\ref{noise_qubits-higher-thermal} and Fig. ~\ref{noise-diagram}. \\

\noindent{\textit{Definitions of $T_1$ and $T_2$} \\
$T_1$ and $T_2$ values for the qubits are generated randomly based on a normal distribution, where $T_1$ averages 50 microseconds with a standard deviation of 10 microseconds, and $T_2$ averages 70 microseconds with a standard deviation of 10 microseconds.\\}

\noindent{\textit{Truncation of $T_2$ Values} \\
It is physically inferred for $T_2$ to be at most twice $T_1$ (since $T_2$ involves both relaxation and other dephasing effects), thus $T_2$ values are truncated to be no more than twice their corresponding $T_1$ values for each qubit.\\}

\noindent{\textit{Quantum Gate and Measurement Times}\\
Instruction Times: These are the durations for each type of quantum operation in nanoseconds. For example, the U2 gate which involves a single X90 pulse takes 50 ns, and the U3 gate which involves two X90 pulses takes 100 ns. Longer operations like measurement and reset are set to 1000 ns (1 microsecond).\\}

\noindent{\textit{Creation of Quantum Errors} \\
Thermal Relaxation Errors: Using the \texttt{thermal\_relaxation\_error(t1, t2, time)} function from Qiskit’s Aer module to simulate quantum noise  \cite{qiskitAer}, thermal relaxation errors are created for each quantum operation based on their durations and the $T_1$ and $T_2$ values of each qubit.}\\

\noindent{\textit{Noise Model Analysis} \\
System Noise for 6 Qubit Hc Hamiltonian: By utilizing the cluster Hamiltonian (Equation  \ref{eq:model-ham}) derived from the one-band and three-band Hubbard models of cuprates \cite{Sheshadri2023}, we can map the hemocyanin Hamiltonian to compute the Heisenberg exchange coupling, $J$, with the aim to examine how the system varies with noise introduced by different parameters, as described in Equation \ref{eq:J}, where site labels $i = 1, 2$ are for impurity 1 and 2 and $i = 3$ for the bath.}\\

\begin{multline}
\hat{H} = \frac{\Delta}{2}*(n_3-n_1-n_2)-t_{pd}\sum_{i,\sigma}(d_{i\sigma}^{{\dagger}}p_{\sigma}+H.c.)+\\
U_d*(n_{1\uparrow}n_{1\downarrow}+n_{2\uparrow}n_{2\downarrow})+U_p*n_{3\uparrow}n_{3\downarrow}
\label{eq:model-ham}
\end{multline}

where \( n_d = d^\dagger_{i\sigma} d_{i\sigma} \) (for \( i=1,2 \)), \( n_{3\sigma} = p^\dagger_\sigma p_\sigma \), and 

\[
n_i = n_{i\uparrow} + n_{i\downarrow} + 2 t_{pd} \left( \frac{1}{J} \right) \left(1 - \frac{1}{2}\right) i\sigma
\]

\noindent{Here, \( d^\dagger_{i\sigma} \) creates a hole with a \( z \) component of spin \( \sigma = \pm 1 \) in the impurities site \( i \) (\( i=1,2 \)), and \( p^\dagger \) creates a hole with a \( z \) component of spin \( \sigma = \pm 1 \) in the bath site. The charge-transfer energy is $\Delta$ and \( U_p \) and \( U_d \) are the on-site Coulomb energies at the bath and impurities sites, respectively, and finally, \( t_{pd} \) is the strength of hopping between neighboring bath and impurities sites.}

\begin{equation}
J = \frac{4 t_{pd}^4}{\Delta^2} \left( \frac{1}{U_d} + \frac{1}{\Delta} \right)
\label{eq:J}
\end{equation}

\begin{equation}
E_{bath}\approx-0.0633 E_h
\label{eq:E-bath}
\end{equation}

\begin{equation}
E_{imp1}\approx-0.2842 E_h
\label{eq:E-imp1}
\end{equation}

\begin{equation}
E_{imp2}\approx-0.2633 E_h
\label{eq:E-imp2}
\end{equation}

\noindent{where $E_{bath}$, $E_{imp1}$ and $E_{imp2}$ are the bath and impurities onsite energies, respectively.}

\begin{equation}
\Delta=E_{bath}-\frac{E_{imp1}+E_{imp2}}{2}\approx0.2104 E_h
\end{equation}

\begin{equation}
U_d\approx0.2934 E_h; t_{pd}\approx0.0578 E_h; U_p\approx0 E_h
\end{equation}

\noindent{Assuming noisy $\Delta'$ and $t_{pd}'$ of the form below:}

\begin{equation}
\Delta' = \Delta + \text{GaussianNoise} \times \text{Amplitude}
\end{equation}

\begin{equation}
t_{pd}' = t_{pd} + \text{GaussianNoise} \times \text{Amplitude}
\end{equation}

\noindent{We anticipate that the effects on $J$ from the noisy terms above will exhibit the following characteristics, Equation \ref{eq:J-t} and Equation \ref{eq:J-delta}, with steps to get to the results.}

\noindent{Using the binomial expansion for \( (t_{pd} + \delta)^4 \):}

\begin{equation}
(t_{pd} + \delta)^4 \approx t_{pd}^4 + 4 t_{pd}^3 \delta
\end{equation}
where \(\delta = \text{GaussianNoise} \times \text{Amplitude}\).

\noindent{For noisy $t_{pd}$ term, the effect on $J$ is:}

\begin{equation}
J' \approx \frac{4 \left(t_{pd}^4 + 4 t_{pd}^3 \delta\right)}{\Delta^2} \left(\frac{1}{U_d} + \frac{1}{\Delta}\right)
\label{eq:J-t}
\end{equation}

\noindent{And for noisy $\Delta$ term, the effect on $J$ is as follows.}

\noindent{Using Taylor expansions for the quadratic term:}

\begin{equation}
(\Delta + \delta)^2 \approx \Delta^2 + 2\Delta \delta
\end{equation}
where \(\delta = \text{GaussianNoise} \times \text{Amplitude}\).

\noindent{For \(\frac{1}{\Delta + \delta}\):}

\begin{equation}
\frac{1}{\Delta + \delta} \approx \frac{1}{\Delta} - \frac{\delta}{\Delta^2}
\end{equation}

\noindent{Finally, we arrive at:}

\begin{equation}
J' \approx \frac{4 t_{pd}^4}{\Delta^2} \left(1 - 2 \frac{\delta}{\Delta}\right) \left( \frac{1}{U_d} + \frac{1}{\Delta} - \frac{\delta}{\Delta^2} \right)
\label{eq:J-delta}
\end{equation}

\noindent{Due to the size of the onsite bath energy, Equation ~\ref{eq:E-bath}, compared to the impurities' energies in Equation ~\ref{eq:E-imp1} and Equation ~\ref{eq:E-imp2}, it is expected that $\Delta$ would not be strongly affected by the noise on the bath site qubits. Taking into account Equation \ref{eq:J} and that $U_d$ would also not be affected by bath site noise, we can conclude that we expect $J$ reaction to bath noise to be dominated by the $t_{pd}$ term. On the other hand, for noise on the impurities, Equation \ref{eq:J-delta} and Equation \ref{eq:J-t} show that competing effects from $\Delta$, $U_d$ and $t_{pd}$ are expected to reduce the overall noise. This effect is depicted in both Fig. \ref{noise_qubits-higher-thermal} and Fig. \ref{noise-diagram}, with noisy bath simulations clearly showing higher error levels for the ground state energy estimations compared to noisy impurity sites.}

In a different approach, in Fig. ~\ref{noise_qubits-depol} and Fig. ~\ref{noise_qubits-depol-corr} we investigated the effects of different levels of depolarisation error (probability of depolarising) applied to all 6 qubits on a simulator with coupling map set to ibm\_Casablanca. In real quantum systems, both amplitude damping ($T1$) and phase damping ($T2$) contribute to the overall error in a quantum state. Depolarising error can be seen as a phenomenological model that encapsulates the net effect of these noises when the detailed behavior of each is not crucial for the high-level simulation. It randomly maps the state of a qubit to the maximally mixed state with a certain probability, thus capturing the overall likelihood of a qubit being in an erroneous state due to any noise. \\

\noindent{{\large Hardware topology}}

\noindent{Quantum computing holds immense opportunity for accessing new ways of computation, but current NISQ devices suffer from many limitations due to the early stage of the field. In classical computation the design of the underlying hardware is abstracted away for the user and operations and measurements do not vary based on the environment of the instantaneous runs, but produce a deterministic outcome. On the contrary, NISQ devices show large differences between the same Hamiltonian measurements with different topology mapping or even for the same mapping and the same hardware but on different days. To investigate the level of sensitivity to the topology of the quantum hardware, we performed ground state energy optimisation on different topology mappings. Due to the strong correlation effects in Hc we expected to observe differences in the results stemming from not total connectivity between the qubits. The topology mappings focused on increasing the physical distance and "quantum" path between the 2 impurity sites, as those are crucial for representing modelled Hc accurately, results are presented in Fig. ~\ref{topology}. \\}

\begin{figure}
\includegraphics[width=0.8\columnwidth]{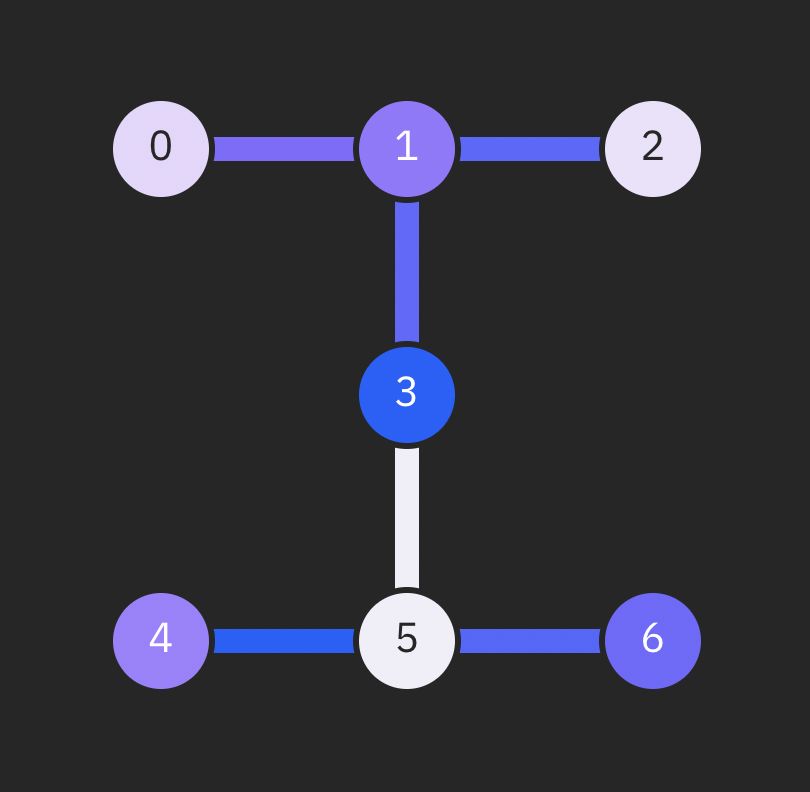}
\caption{Illustration of the coupling map of IBM\_Oslo quantum hardware, delineating which qubits have direct connectivity with each other. Qubits that are not directly connected are also indicated, providing a comprehensive view of the inter-qubit relationships within this quantum hardware configuration.}
\label{hardware-draw}
\end{figure}

\begin{figure}
\includegraphics[width=1\columnwidth]{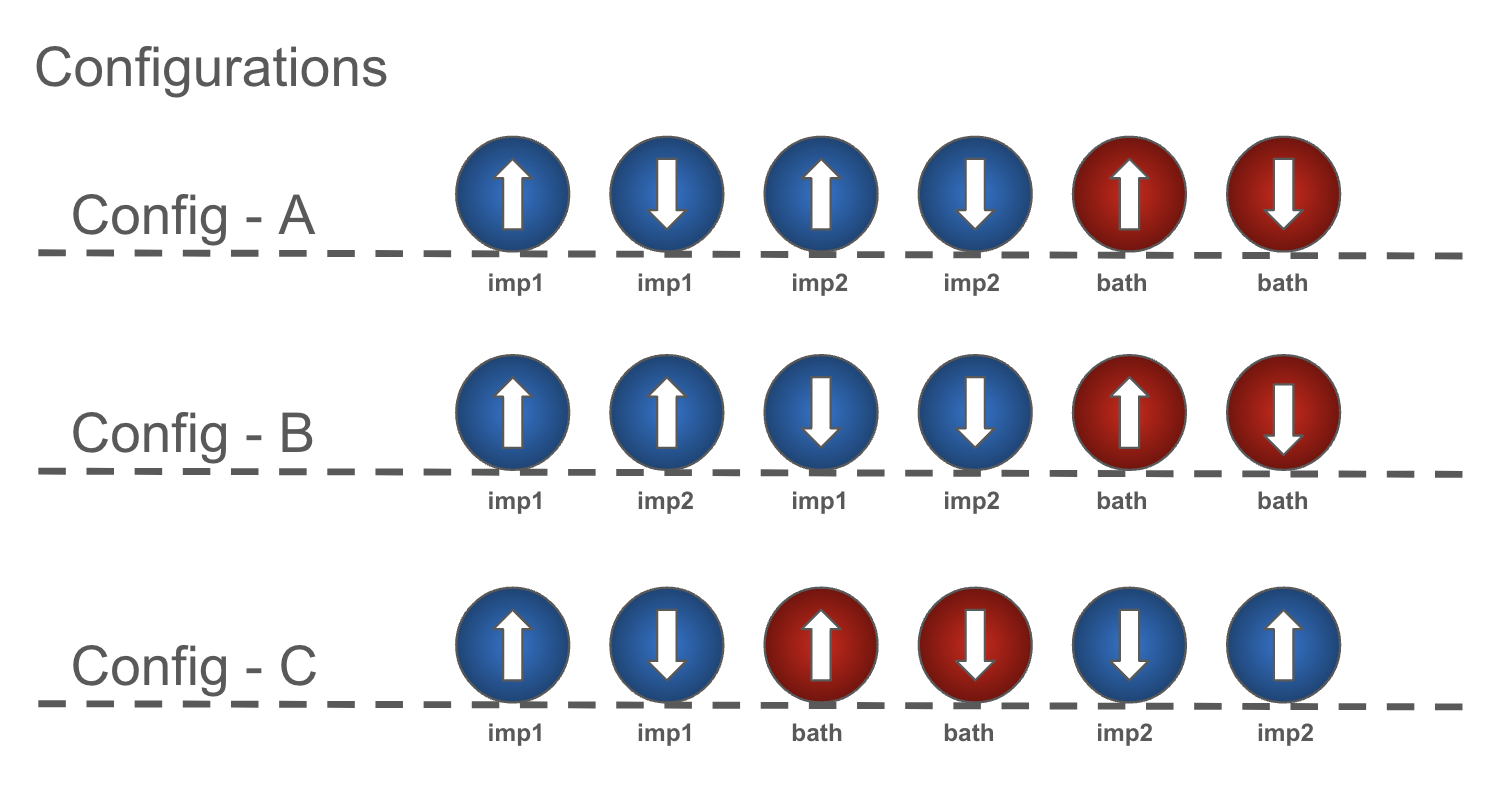}
\caption{Qubit mappings of Anderson impurity model of Hc molecule with two impurities (blue spheres) and one bath site (red spheres). The configurations mark different ordering of the sites onto the underlying hardware with the same coupling map, reference Fig. ~\ref{hardware-draw}.}
\label{config-draw}
\end{figure}

\begin{figure}
\includegraphics[width=1\columnwidth]{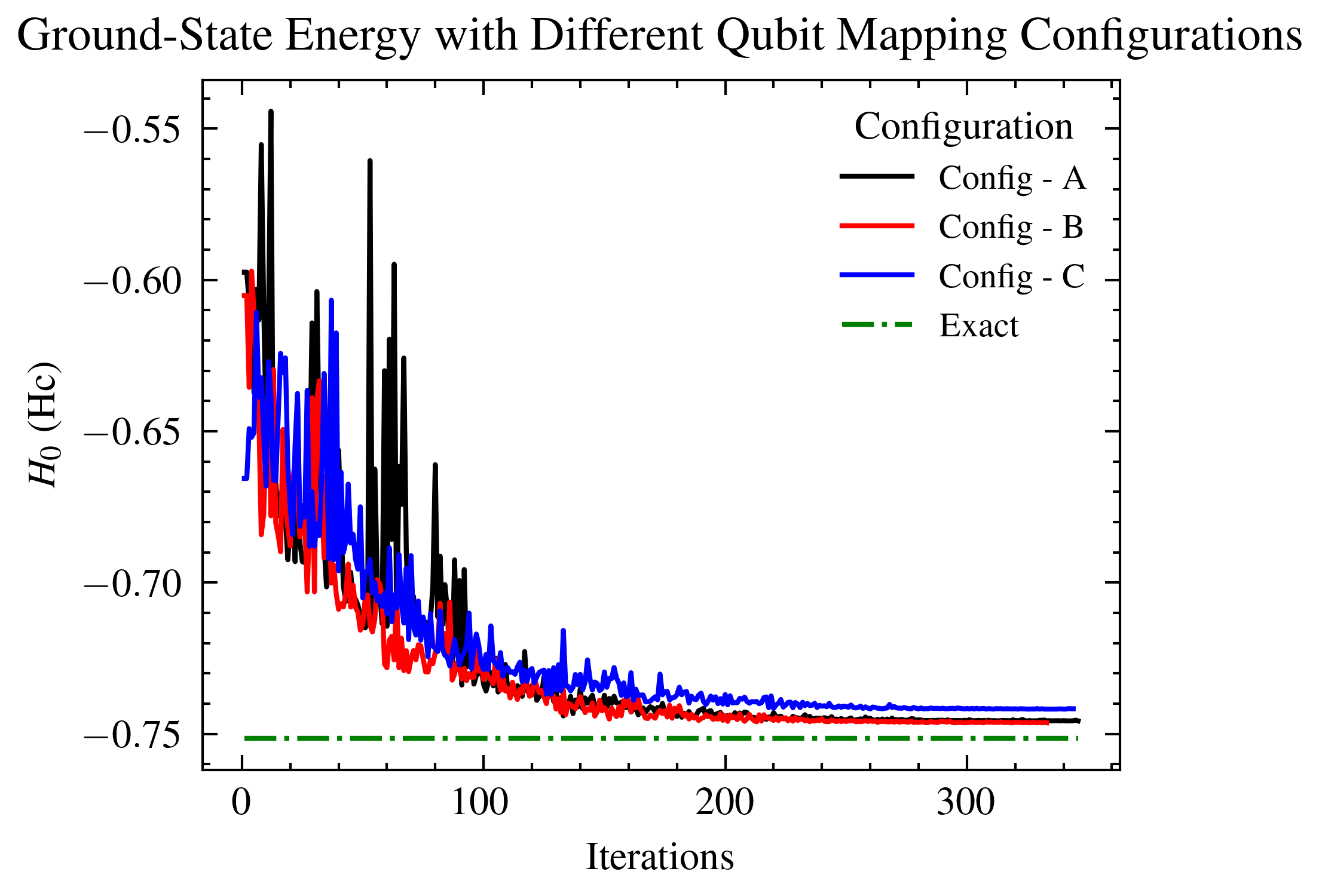}
\caption{Comparison of ground-state energy estimates in $E_h$ derived from different topological mappings of a 6-qubit hemocyanin Hamiltonian onto the qubits. The labels of the configurations correlate to the mappings illustrated in Fig. ~\ref{config-draw}, and each simulation employed a Generalised UCCSD ansatz with the same coupling map as shown in Fig. ~\ref{hardware-draw}. Notably, $Config - C$, which positions the two impurities furthest apart, exhibits the highest error, as anticipated due to the limited connectivity between the most correlated sites of the Hamiltonian. The other two configurations yield similar results, demonstrating the impact of qubit arrangement on simulation accuracy.}
\label{topology}
\end{figure}

\noindent{{\large Onsite potential}}

\noindent{The complex Hc superexchange Cu-O-Cu pathway has been investigated extensively due to its interesting character. Variation in the distance between the mean position of the two copper atoms and the mean position of the two oxygen atoms has shown a singlet-to-triplet transition that occurs at $R=0.6 A$, where $R = |\frac{1}{2}*(r_{CuA}+r_{CuB})-\frac{1}{2}*(r_{O1}+r_{O2})|$ \cite{Cedric-hemocyanin}. To investigate this transition we varied the onsite potential $U$ which accounts for a similar calculation as varying $R$. Fig. ~\ref{onsite} shows the ground state convergence of VQE runs for 3 values of the onsite potential. \\}

\begin{figure}
\includegraphics[width=1\columnwidth]{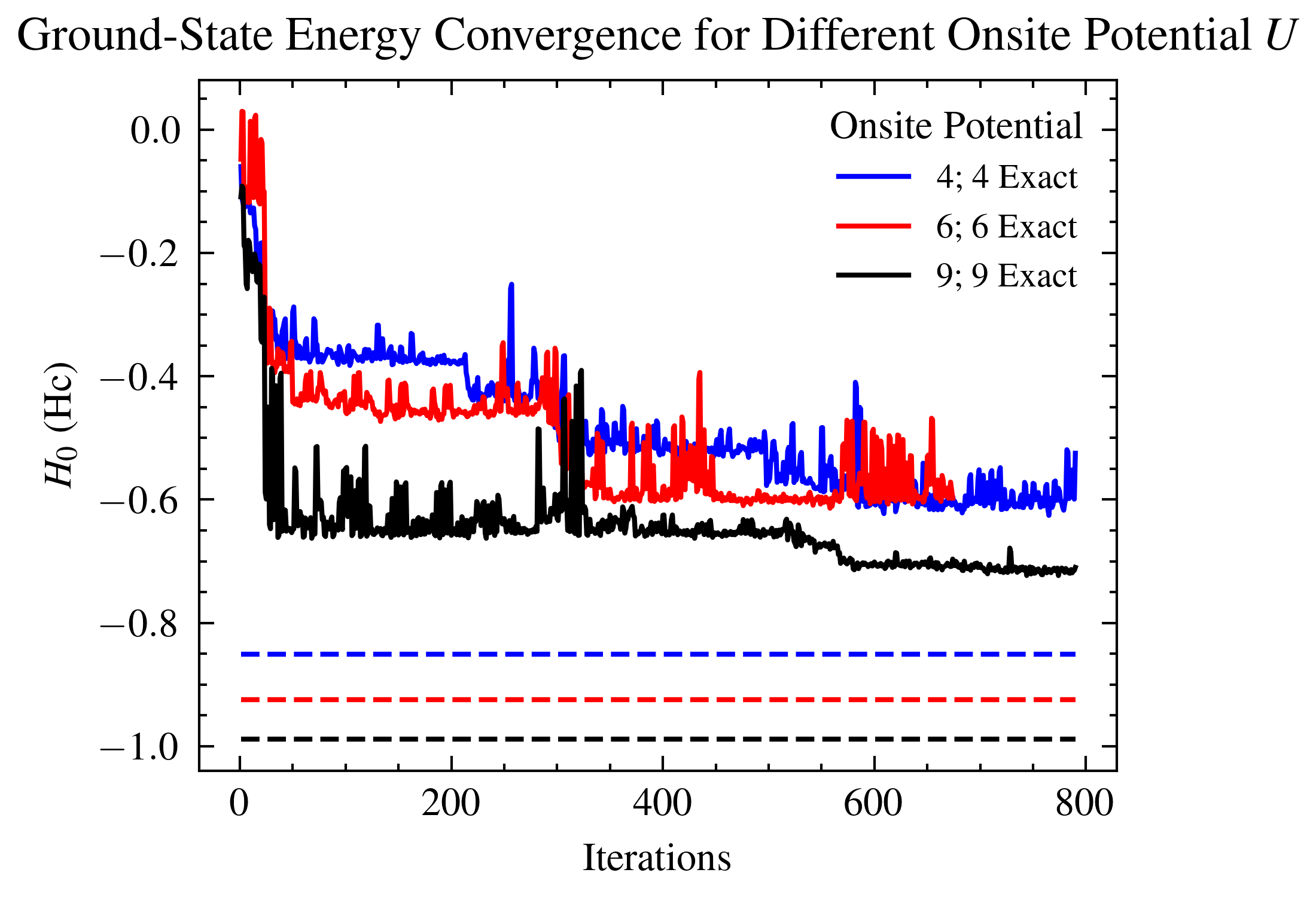}
\caption{Ground-state energy estimates in $E_h$ of Hc 14-qubit mapped Hamiltonian for different values of the onsite potential $U$ in $eV$. VQE simulation setup using Generalised UCCSD ansatz. As expected largest error in the graph is the $U=6 eV$ simulation, which is the suggested point of singlet-triplet transition from ~\cite{Cedric-hemocyanin}.}
\label{onsite}
\end{figure}

\noindent{{\large Correlation}}

\noindent{Conventional many body methods struggle to correctly predict Hc properties due to the complexity of the superexchange pathway between the Cu d-orbitals and intermediate O p-orbitals. The AIM model showed a strong correlation between the 2 impurity sites that is difficult to represent on a classical computer simulation, but trivial to map onto qubit interaction terms. Table ~\ref{correlation_t} represents a comparison between spin-spin correlation measurement for different VQE setups that outlines the sensitivity to the topology mapping of the Hamiltonian and to the form of the ansatz. Furthermore, Fig. ~\ref{noise_qubits-depol-corr} shows the effect of same levels of depolarisation noise on both ground state energy estimation and impurity spin-spin correlation to assess the stability to noise when measuring properties other than the energy.} 

\begin{table}
\centering
\caption{\textbf{Hemocyanin AIM impurities Spin-Spin correlation measurements 6 Qubits}}
\label{correlation_t}
\begin{tabular}{|c|c|c|}
 \hline
 \multicolumn{3}{|c|}{Spin-Spin Correlation} \\
 \hline
 Topology & VQE Setup & \(\langle S_z[i] * S_z[j] \rangle\) \\
 \hline
  & Exact & -0.307     \\
 Config-A & Spin-conserved UCCSD & -0.314 \\
 Config-A & UCCS & -0.248 \\
 Config-A & Generalised UCCSD & -0.169 \\
 Config-B & Generalised UCCSD & -0.018 \\
 Config-C & Generalised UCCSD & -0.078 \\
  \hline
\end{tabular}
\end{table}

\begin{table}
\centering
\caption{\textbf{The simulation of hemocyanin using the Generalized UCCSD ansatz with the $Config - A$ mapping resulted in energy estimates closest to the exact value, based on the following operator counts.
}}
\label{operators}
\begin{tabular}{|c|c|}
 \hline
 \multicolumn{2}{|c|}{Generalised UCCSD $Config - A$} \\
\hline
Operator & Count \\
\hline
U2 Gate & 288 \\
\hline
CNOT Gate (Controlled-NOT) & 280 \\
\hline
U1 Gate & 184 \\
\hline
U Gate & 4 \\
\hline
\end{tabular}
\end{table}

\section{Discussion} 
Here we investigated the application of VQE method for characterisation of a prototypal anti-tumor vaccine metalloprotein (Hc) for the first time on quantum hardware and simulator, reference Fig. ~\ref{ground_energy_hemocyanin}. Varying the configuration of the VQE setup allowed us to understand in depth the dynamics of the simulation and the driving sources of error. Such analysis proves to be promising for understanding complex entangled systems on quantum devices and their best representation. With the steep improvement of tackling noise on these devices and the growth of the number of qubits and their connectivity in available quantum hardware (Fig. ~\ref{quantum_experiments_plot}) soon it could be possible to map larger highly correlated systems directly onto qubits for characterisation. Notable example of related work is the recently developed end-to-end quantum simulation workflow by Microsoft Azure Quantum for chemical systems \cite{AzureQuantum}. Leveraging entangled qubits to model highly correlated electron systems on quantum hardware offers a substantial enhancement over traditional computing methods. This is primarily because the underlying principles of quantum mechanics govern both the methodology and the systems under investigation, ensuring a more natural and accurate representation.

As illustrated in Fig. ~\ref{ground_energy_hemocyanin}, the noiseless simulator's VQE run converges to the correct result in fewer than 150 iterations, demonstrating the potential efficacy of simulating quantum phenomena accurately. Conversely, the noisy simulator, incorporating the full spectrum of hardware noise reproduced for the same hardware, struggled to maintain a consistent energy minimization pattern. This challenge primarily stems from the absence of error mitigation techniques, which are crucial in counteracting the inherent noise present on quantum hardware. Furthermore, Fig. ~\ref{ground_energy_hemocyanin} also highlights the impact of error mitigation in quantum computing simulations, where the hardware execution exhibits a marked convergence, characterized by a significant energy reduction within the first 100 iterations followed by a more gradual decline. This underscores the importance of error mitigation in enhancing the accuracy and reliability of quantum simulations.

To evaluate the effectiveness of different VQE setups, we conducted simulations using various ansatz types, as shown in Fig. ~\ref{ansatz}. The outcomes demonstrated that both the generalized UCCS and generalized UCCSD ansatze converged to the exact value, with UCCS achieving convergence in fewer iterations (under 50), whereas UCCSD required between 100 to 150 iterations. The latter involves second-order excitations, necessitating a larger pool of ansatz operators and longer circuits, which introduce more noise. However, it captures a broader range of the underlying physical processes. Under conditions of moderate noise, UCCSD is likely to yield superior results. Operators count for the Generalised UCCSD ansatz with $Config - A$ can be found in Table ~\ref{operators}.

Table ~\ref{correlation_t} presents the spin-spin correlation of impurities measured via VQE simulations, indicating the most accurate results were obtained with the Spin Conserved UCCSD ansatz, which adheres closely to the physical characteristics of the system. In Fig. ~\ref{ansatz}, the simulations for the ground state energy using Spin Conserved UCCSD and EfficientSU2 HEA ansatze both converge to a barren plateau level \cite{barren}, despite the former closely mirroring true-value impurity correlations. These results underscore the enhancements in system representation when restricting to physically justifiable excitations, though they also highlight potential limitations due to the limited selection of operators in the pool. Future investigations might explore the spin-conserved UCC with third-order excitations to potentially overcome these challenges.

In order to probe the stability of the simulation, we applied $T1/T2$ thermal relaxation noise to different sets of qubits and observed the effect on the ground state energy, reference Fig. ~\ref{noise_qubits-higher-thermal}, and further investigated the effect of depolarisation error applied to the whole system, Fig ~\ref{noise_qubits-depol}. Due to the lack of error mitigation, the addition of noise proves to be critical for the simulations. Notably, introducing noise to the qubits representing impurity 1 or impurity 2, as expected from their correlation, yielded similar effects on the ground state energy (refer to Fig. ~\ref{noise_qubits-higher-thermal}). Conversely, simulations involving noise at the bath site exhibited a less stable trajectory. These findings underscore the critical role of inter-impurity correlations in the simulation;

Furthermore, investigating the effect of depolarising error on the system, Fig. ~\ref{noise_qubits-depol} and Fig. ~\ref{noise_qubits-depol-corr}, highlighted a strong quadratic dependence of the ground state energy deviation versus the depolarising error rate. A slope of approximately $2.01$ in a log-log plot of ground-state energy deviation versus depolarising error rate suggests a specific power-law relationship between the two variables. This means that the ground-state energy deviation varies as the square of the depolarising error rate, defined mathematically as:

\begin{equation}
\log(\text{Energy Deviation}) \propto 2.01 \cdot \log(\text{Depolarising Error})
\end{equation}

\begin{equation}
\text{Energy Deviation} \propto (\text{Depolarising Error})^{2.01}
\end{equation}

A slope of approximately $2$ highlights a high sensitivity of the quantum simulation or computation to depolarising errors. Small increases in error rate cause significant deviations in the calculated ground-state energy, which can severely impact the accuracy and reliability of the measurements. The quadratic relationship emphasizes the compounding effect of errors in quantum systems. This is particularly critical in quantum computations, where maintaining coherence and precision is essential for obtaining correct results. Knowing that the relationship is quadratic allows for more precise modeling and predictions regarding the behavior of a quantum system under noise. This can aid in the design and scaling of quantum algorithms by setting thresholds for acceptable error rates to achieve desired accuracies. Additionally, we present Fig. ~\ref{noise_qubits-depol-corr}, which illustrates the variations in ground state energy and spin-spin correlation deviations across different levels of depolarization noise. The percentage error in energy relative to the spin-spin correlation highlights the significant sensitivity of correlation measurements to noise. For instance, $5\%$ error of the noisy to noiseless simulation result for the ground state energy is equivalent to up to $80\%$ error in the correlation estimate. This is an important point to note when considering characterisation of materials properties with quantum hardware.

To evaluate the VQE method, one effective approach is to vary the mappings of the impurity and bath sites onto the qubits. Given the known spin-spin correlations between impurities, it is expected that positioning their corresponding qubits further apart would lead to less accurate results. Fig. ~\ref{config-draw} illustrates the three different configuration setups used in the experiments, with $Config - C$ having the impurities' qubits at the greatest distance from each other. The results, presented in Fig. ~\ref{topology}, confirm this hypothesis: the $Config - C$ simulation converges more slowly to a higher estimate of the ground state energy, primarily due to noise effects associated with entangling more distantly located qubits.

Each experiment was conducted using the same coupling map to maintain consistent connectivity between qubits, as dictated by the hardware used in the simulations. Over the years, the connectivity of qubits in superconducting quantum hardware has evolved but remains limited, highlighting the significance of qubit mapping in quantum simulations. Our findings emphasize that the physical layout of the hardware is crucial and should not be overlooked, as it significantly affects the quality of the simulations. Unlike classical computing environments, the layout in quantum computing setups is not yet sufficiently abstracted to ensure that results are independent of qubit ordering. For an in-depth examination of advancements in qubit connectivity within quantum computing hardware, comprehensive analyses are available through recent scholarly reviews \cite{Preskill2018, Arute2019} and industry developments, particularly from leading entities like IBM \cite{IBMQuantum} and Google \cite{GoogleQuantumAI}. These sources collectively point at the progressive enhancements in qubit design and deployment, significantly influencing the field's evolution.

In this study, we introduce the first application of a VQE simulation on an AIM model for a highly correlated electron system (Hc) utilizing both quantum hardware and simulators. Adjusting the parameters within the VQE setup facilitated not only the validation of the simulation but also the optimization of the configuration to achieve the best results. The incorporation of noise models, as depicted in Fig. ~\ref{noise_qubits-higher-thermal} and Fig. ~\ref{noise_qubits-depol}, along with the strategic alteration of qubit mappings for the impurities' sites, as shown in Fig. ~\ref{topology}, confirmed the alignment between the anticipated trends and the actual convergence to the ground state energy. While the VQE method on contemporary quantum machines is challenged by noise effects, it still showcases a significant capability to surpass traditional quantum chemistry models on classical devices. This advantage is increasingly amplified through advancements in error mitigation strategies and the ongoing reduction of hardware noise. The Hamiltonians utilized in this study are openly accessible to the community for benchmarking and evaluating enhancements in hardware efficiency.

\section*{Data Availability}

The data supporting the findings of this study are available from the corresponding authors upon reasonable request.

\quad

\section*{Code Availability}

The code developed throughout this investigation is available at \href{https://github.com/elenachachkarova/hemocyanin-vqe}{GitHub Repository}.

\section*{Author Contribution}

The calculations have been carried out by E.C., T.T. and C.W.. All of the authors wrote this paper. This investigation was outlined by C.W.. The published version of the paper has been read and reviewed by all authors.

\section*{Funding}

E.C. and T.T. are self-funded students. Y.W. is grateful for funding from the China Scholarship Council.

\section*{Supplementary information}
Supplementary information \href{run:./supp.pdf}{here}. 

\section*{Acknowledgment}

The authors are thankful for valuable discussions with Professor Joe Bhaseen and Maria Reboredo Prado. The research was conducted on free resources supplied by IBM Quantum platform \cite{IBM} and Quantinuum \cite{quantinuum}. This research was lead by Dr Cedric Weber.

\bibliographystyle{apsrev4-2}
\bibliography{apssamp}

\end{document}


\onecolumngrid

\section{Supplementary Information}

Presented below are the constructed Hamiltonians used in the VQE solver in its transformed Pauli string form, corresponding to the number of qubits and value of U.

\begin{lstlisting}
6 Qubit
(-0.4640485702054111+0j) [] +
(-0.0332912087832737+0j) [X0 Z1 Z2 Z3 X4] +
(-0.0332912087832737+0j) [Y0 Z1 Z2 Z3 Y4] +
(0.06874150000000001+0j) [Z0] +
(0.07335+0j) [Z0 Z1] +
(-0.0332912087832737+0j) [X1 Z2 Z3 Z4 X5] +
(-0.0332912087832737+0j) [Y1 Z2 Z3 Z4 Y5] +
(0.06874150000000001+0j) [Z1] +
(0.0245514029353716+0j) [X2 Z3 X4] +
(0.0245514029353716+0j) [Y2 Z3 Y4] +
(0.058285350000000014+0j) [Z2] +
(0.07335+0j) [Z2 Z3] +
(0.0245514029353716+0j) [X3 Z4 X5] +
(0.0245514029353716+0j) [Y3 Z4 Y5] +
(0.058285350000000014+0j) [Z3] +
(0.0316474351027055+0j) [Z4] +
(0.0316474351027055+0j) [Z5]

8 Qubit
(-0.41926934879169153+0j) [] +
(0.0169273012830259+0j) [X0 Z1 Z2 Z3 X4] +
(-0.0124834455965616+0j) [X0 Z1 Z2 Z3 Z4 Z5 X6] +
(0.0169273012830259+0j) [Y0 Z1 Z2 Z3 Y4] +
(-0.0124834455965616+0j) [Y0 Z1 Z2 Z3 Z4 Z5 Y6] +
(0.06874150000000001+0j) [Z0] +
(0.07335+0j) [Z0 Z1] +
(0.0169273012830259+0j) [X1 Z2 Z3 Z4 X5] +
(-0.0124834455965616+0j) [X1 Z2 Z3 Z4 Z5 Z6 X7] +
(0.0169273012830259+0j) [Y1 Z2 Z3 Z4 Y5] +
(-0.0124834455965616+0j) [Y1 Z2 Z3 Z4 Z5 Z6 Y7] +
(0.06874150000000001+0j) [Z1] +
(-0.0286665490477019+0j) [X2 Z3 X4] +
(0.02114083628193125+0j) [X2 Z3 Z4 Z5 X6] +
(-0.0286665490477019+0j) [Y2 Z3 Y4] +
(0.02114083628193125+0j) [Y2 Z3 Z4 Z5 Y6] +
(0.058285350000000014+0j) [Z2] +
(0.07335+0j) [Z2 Z3] +
(-0.0286665490477019+0j) [X3 Z4 X5] +
(0.02114083628193125+0j) [X3 Z4 Z5 Z6 X7] +
(-0.0286665490477019+0j) [Y3 Z4 Y5] +
(0.02114083628193125+0j) [Y3 Z4 Z5 Z6 Y7] +
(0.058285350000000014+0j) [Z3] +
(0.0236589565634185+0j) [X4 Z5 X6] +
(0.0236589565634185+0j) [Y4 Z5 Y6] +
(-0.00841924169912505+0j) [Z4] +
(0.0236589565634185+0j) [X5 Z6 X7] +
(0.0236589565634185+0j) [Y5 Z6 Y7] +
(-0.00841924169912505+0j) [Z5] +
(0.0176770660949708+0j) [Z6] +
(0.0176770660949708+0j) [Z7]

14 Qubit
(-0.9203822755004079+0j) [] +
(-0.0102330968552639+0j) [X0 Z1 Z2 Z3 X4] +
(-0.0058645309558+0j) [X0 Z1 Z2 Z3 Z4 Z5 X6] +
(-0.0232723792309368+0j) [X0 Z1 Z2 Z3 Z4 Z5 Z6 Z7 X8] +
(0.0285112773771463+0j) [X0 Z1 Z2 Z3 Z4 Z5 Z6 Z7 Z8 Z9 X10] +
(0.03368107911672325+0j) [X0 Z1 Z2 Z3 Z4 Z5 Z6 Z7 Z8 Z9 Z10 Z11 X12] +
(-0.0102330968552639+0j) [Y0 Z1 Z2 Z3 Y4] +
(-0.0058645309558+0j) [Y0 Z1 Z2 Z3 Z4 Z5 Y6] +
(-0.0232723792309368+0j) [Y0 Z1 Z2 Z3 Z4 Z5 Z6 Z7 Y8] +
(0.0285112773771463+0j) [Y0 Z1 Z2 Z3 Z4 Z5 Z6 Z7 Z8 Z9 Y10] +
(0.03368107911672325+0j) [Y0 Z1 Z2 Z3 Z4 Z5 Z6 Z7 Z8 Z9 Z10 Z11 Y12] +
(0.06874150000000001+0j) [Z0] +
(0.07335+0j) [Z0 Z1] +
(-0.0102330968552639+0j) [X1 Z2 Z3 Z4 X5] +
(-0.0058645309558+0j) [X1 Z2 Z3 Z4 Z5 Z6 X7] +
(-0.0232723792309368+0j) [X1 Z2 Z3 Z4 Z5 Z6 Z7 Z8 X9] +
(0.0285112773771463+0j) [X1 Z2 Z3 Z4 Z5 Z6 Z7 Z8 Z9 Z10 X11] +
(0.03368107911672325+0j) [X1 Z2 Z3 Z4 Z5 Z6 Z7 Z8 Z9 Z10 Z11 Z12 X13] +
(-0.0102330968552639+0j) [Y1 Z2 Z3 Z4 Y5] +
(-0.0058645309558+0j) [Y1 Z2 Z3 Z4 Z5 Z6 Y7] +
(-0.0232723792309368+0j) [Y1 Z2 Z3 Z4 Z5 Z6 Z7 Z8 Y9] +
(0.0285112773771463+0j) [Y1 Z2 Z3 Z4 Z5 Z6 Z7 Z8 Z9 Z10 Y11] +
(0.03368107911672325+0j) [Y1 Z2 Z3 Z4 Z5 Z6 Z7 Z8 Z9 Z10 Z11 Z12 Y13] +
(0.06874150000000001+0j) [Z1] +
(-0.01395767436142885+0j) [X2 Z3 X4] +
(0.02042556305865555+0j) [X2 Z3 Z4 Z5 X6] +
(-0.0342747841006893+0j) [X2 Z3 Z4 Z5 Z6 Z7 X8] +
(-0.0194319517229672+0j) [X2 Z3 Z4 Z5 Z6 Z7 Z8 Z9 X10] +
(-0.0051669611836042+0j) [X2 Z3 Z4 Z5 Z6 Z7 Z8 Z9 Z10 Z11 X12] +
(-0.01395767436142885+0j) [Y2 Z3 Y4] +
(0.02042556305865555+0j) [Y2 Z3 Z4 Z5 Y6] +
(-0.0342747841006893+0j) [Y2 Z3 Z4 Z5 Z6 Z7 Y8] +
(-0.0194319517229672+0j) [Y2 Z3 Z4 Z5 Z6 Z7 Z8 Z9 Y10] +
(-0.0051669611836042+0j) [Y2 Z3 Z4 Z5 Z6 Z7 Z8 Z9 Z10 Z11 Y12] +
(0.058285350000000014+0j) [Z2] +
(0.07335+0j) [Z2 Z3] +
(-0.01395767436142885+0j) [X3 Z4 X5] +
(0.02042556305865555+0j) [X3 Z4 Z5 Z6 X7] +
(-0.0342747841006893+0j) [X3 Z4 Z5 Z6 Z7 Z8 X9] +
(-0.0194319517229672+0j) [X3 Z4 Z5 Z6 Z7 Z8 Z9 Z10 X11] +
(-0.0051669611836042+0j) [X3 Z4 Z5 Z6 Z7 Z8 Z9 Z10 Z11 Z12 X13] +
(-0.01395767436142885+0j) [Y3 Z4 Y5] +
(0.02042556305865555+0j) [Y3 Z4 Z5 Z6 Y7] +
(-0.0342747841006893+0j) [Y3 Z4 Z5 Z6 Z7 Z8 Y9] +
(-0.0194319517229672+0j) [Y3 Z4 Z5 Z6 Z7 Z8 Z9 Z10 Y11] +
(-0.0051669611836042+0j) [Y3 Z4 Z5 Z6 Z7 Z8 Z9 Z10 Z11 Z12 Y13] +
(0.058285350000000014+0j) [Z3] +
(0.00800357635717085+0j) [X4 Z5 X6] +
(-0.04725338824796345+0j) [X4 Z5 Z6 Z7 X8] +
(0.00696745171857815+0j) [X4 Z5 Z6 Z7 Z8 Z9 X10] +
(0.0036965854307191+0j) [X4 Z5 Z6 Z7 Z8 Z9 Z10 Z11 X12] +
(0.00800357635717085+0j) [Y4 Z5 Y6] +
(-0.04725338824796345+0j) [Y4 Z5 Z6 Z7 Y8] +
(0.00696745171857815+0j) [Y4 Z5 Z6 Z7 Z8 Z9 Y10] +
(0.0036965854307191+0j) [Y4 Z5 Z6 Z7 Z8 Z9 Z10 Z11 Y12] +
(0.05178138173911655+0j) [Z4] +
(0.00800357635717085+0j) [X5 Z6 X7] +
(-0.04725338824796345+0j) [X5 Z6 Z7 Z8 X9] +
(0.00696745171857815+0j) [X5 Z6 Z7 Z8 Z9 Z10 X11] +
(0.0036965854307191+0j) [X5 Z6 Z7 Z8 Z9 Z10 Z11 Z12 X13] +
(0.00800357635717085+0j) [Y5 Z6 Y7] +
(-0.04725338824796345+0j) [Y5 Z6 Z7 Z8 Y9] +
(0.00696745171857815+0j) [Y5 Z6 Z7 Z8 Z9 Z10 Y11] +
(0.0036965854307191+0j) [Y5 Z6 Z7 Z8 Z9 Z10 Z11 Z12 Y13] +
(0.05178138173911655+0j) [Z5] +
(0.0043850975092616+0j) [X6 Z7 X8] +
(-0.01487110793047115+0j) [X6 Z7 Z8 Z9 X10] +
(-0.02364586870809315+0j) [X6 Z7 Z8 Z9 Z10 Z11 X12] +
(0.0043850975092616+0j) [Y6 Z7 Y8] +
(-0.01487110793047115+0j) [Y6 Z7 Z8 Z9 Y10] +
(-0.02364586870809315+0j) [Y6 Z7 Z8 Z9 Z10 Z11 Y12] +
(0.03514668312536425+0j) [Z6] +
(0.0043850975092616+0j) [X7 Z8 X9] +
(-0.01487110793047115+0j) [X7 Z8 Z9 Z10 X11] +
(-0.02364586870809315+0j) [X7 Z8 Z9 Z10 Z11 Z12 X13] +
(0.0043850975092616+0j) [Y7 Z8 Y9] +
(-0.01487110793047115+0j) [Y7 Z8 Z9 Z10 Y11] +
(-0.02364586870809315+0j) [Y7 Z8 Z9 Z10 Z11 Z12 Y13] +
(0.03514668312536425+0j) [Z7] +
(-0.01037904738535395+0j) [X8 Z9 X10] +
(0.00386000942055545+0j) [X8 Z9 Z10 Z11 X12] +
(-0.01037904738535395+0j) [Y8 Z9 Y10] +
(0.00386000942055545+0j) [Y8 Z9 Z10 Z11 Y12] +
(0.07645310066235694+0j) [Z8] +
(-0.01037904738535395+0j) [X9 Z10 X11] +
(0.00386000942055545+0j) [X9 Z10 Z11 Z12 X13] +
(-0.01037904738535395+0j) [Y9 Z10 Y11] +
(0.00386000942055545+0j) [Y9 Z10 Z11 Z12 Y13] +
(0.07645310066235694+0j) [Z9] +
(-0.0472194011845761+0j) [X10 Z11 X12] +
(-0.0472194011845761+0j) [Y10 Z11 Y12] +
(0.0472879363412088+0j) [Z10] +
(-0.0472194011845761+0j) [X11 Z12 X13] +
(-0.0472194011845761+0j) [Y11 Z12 Y13] +
(0.0472879363412088+0j) [Z11] +
(0.0491451858821573+0j) [Z12] +
(0.0491451858821573+0j) [Z13]
\end{lstlisting}

\begin{figure}
\includegraphics[width=1\columnwidth]{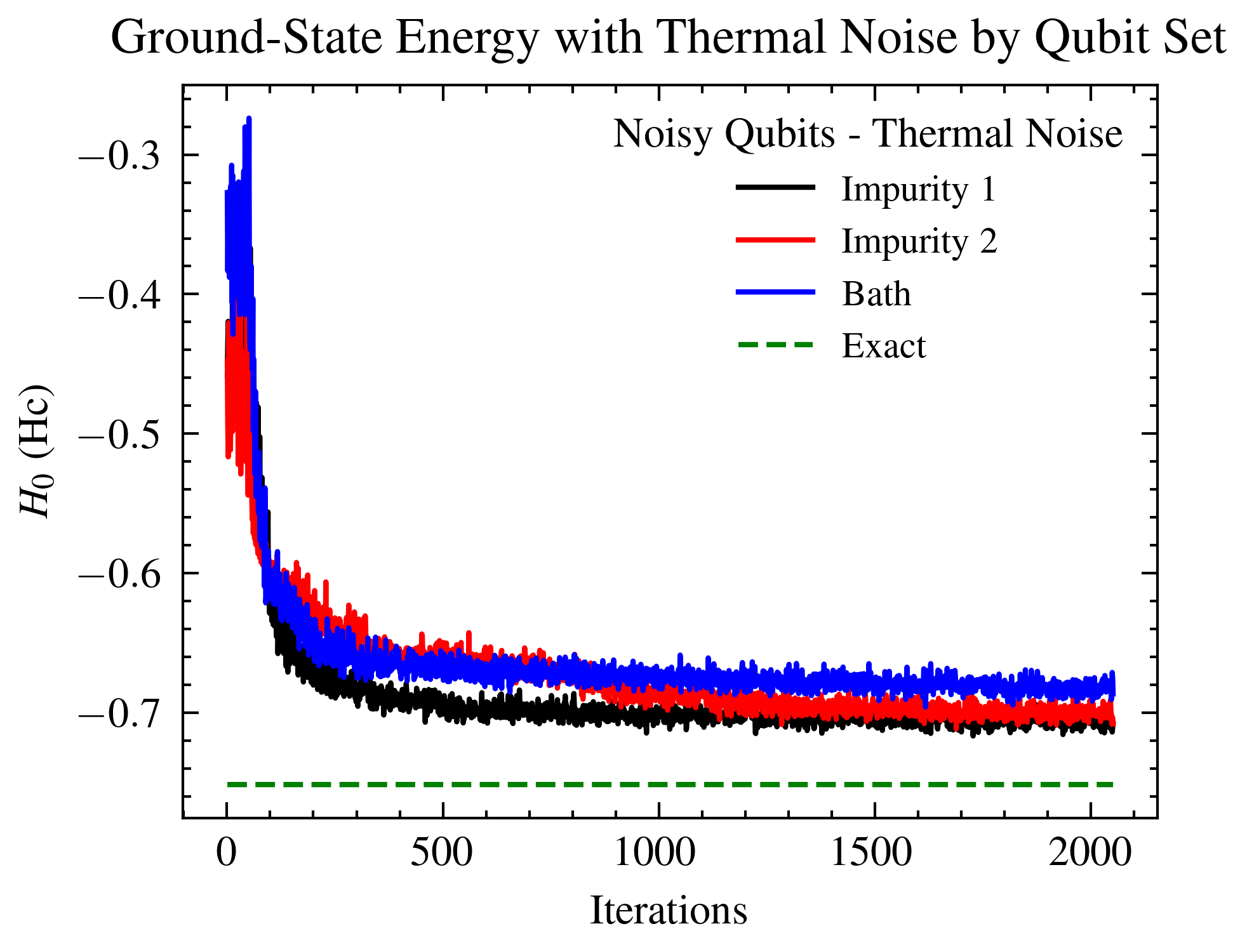}
\caption{Comparison of ground-state energy estimates from various 6-qubit VQE simulation setups utilizing the Generalised UCCSD ansatz, Hamiltonian $Config - A$, and specific VQE parameters. The simulations were subjected to \(T_1/T_2\) thermal relaxation noise of half the level presented in the paper and also selectively applied to three different sets of qubits: Impurity \(1\) (qubits \(0\) and \(1\)), Impurity \(2\) (qubits \(2\) and \(3\)), and the Bath (qubits \(4\) and \(5\)). Consistent with expectations due to the correlation between Impurity \(1\) and Impurity \(2\) sites, noise impacting either set results in similar values for the ground state energy. Conversely, noise affecting the Bath sites demonstrates more significant detrimental effects on the system's performance.}
\label{noise_qubits}
\end{figure}

\begin{figure}
\includegraphics[width=1\columnwidth]{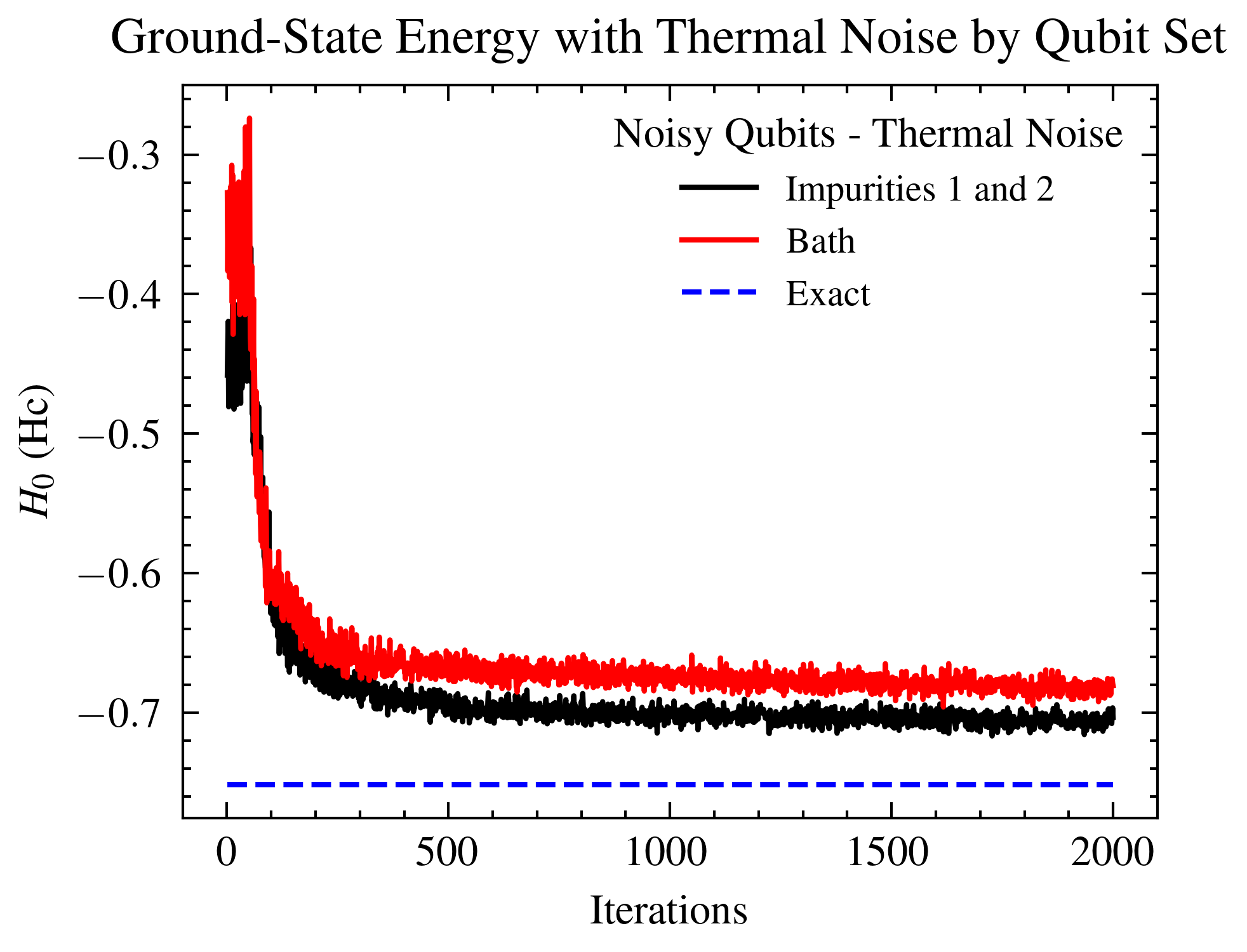}
\caption{Comparison of ground-state energy estimates from various 6-qubit VQE simulation setups utilizing the Generalised UCCSD ansatz, Hamiltonian $Config - A$, and specific VQE parameters. The simulations were subjected to \(T_1/T_2\) thermal relaxation noise, selectively applied to three different sets of qubits: Impurity \(1\) (qubits \(0\) and \(1\)), Impurity \(2\) (qubits \(2\) and \(3\)), and the Bath (qubits \(4\) and \(5\)). Consistent with expectations due to the correlation between Impurity \(1\) and Impurity \(2\) sites, noise impacting either set results in similar values for the ground state energy. Conversely, noise affecting the Bath sites demonstrates more significant detrimental effects on the system's performance.}
\label{noise_qubits-thermal-impurities}
\end{figure}